\documentclass[a4paper,useAMS,usenatbib]{mnras}
\pdfoutput=1

\usepackage{graphicx}
\usepackage{setspace}
\usepackage{natbib}
\usepackage{color}
\usepackage{amsmath,amssymb}
\usepackage{times}
\usepackage{hyperref}[]

\newcommand\name{{\it Aurora}}
\newcommand\namesp{{\it Aurora}\ }

\title[Milky Way's turbulent youth revealed by APOGEE+{\it Gaia} data]{From dawn till disk: Milky Way's turbulent youth revealed by the APOGEE+{\it Gaia} data}

\author[Belokurov \& Kravtsov]{Vasily  Belokurov$^{1,2}$\thanks{E-mail:vasily@ast.cam.ac.uk} and Andrey Kravtsov$^{3,4,5}$\thanks{E-mail:kravtsov@uchicago.edu}\\
  $^1$Institute of Astronomy, Madingley Rd, Cambridge, CB3 0HA, UK\\ 
  $^2$Center for Computational Astrophysics, Flatiron Institute, 162 5th Avenue, New York, NY 10010, USA\\
  $^3$Department of Astronomy and Astrophysics, The University of Chicago, Chicago, IL 60637 USA\\
  $^4$Kavli Institute for Cosmological Physics, The University of Chicago, Chicago, IL 60637 USA\\
  $^5$Enrico Fermi Institute, The University of Chicago, Chicago, IL 60637}

\begin{document}


\maketitle

\label{firstpage}

\begin{abstract}
We use accurate estimates of aluminium abundance provided as part of the APOGEE Data Release 17 and {\it Gaia} Early Data Release 3 astrometry to select a highly pure sample of stars with metallicity $-1.5\lesssim {\rm [Fe/H]}\lesssim 0.5$ born {\it in-situ} in the Milky Way proper. We show that the low-metallicity ([Fe/H]$\lesssim -1.3$) in-situ component that we dub \namesp is kinematically hot with an approximately isotropic velocity ellipsoid and a modest net rotation. \namesp stars exhibit large scatter in metallicity and in a number of element abundance ratios.  The median tangential velocity of the in-situ stars increases sharply with increasing metallicity between [Fe/H]$=-1.3$ and $-0.9$, the transition that we call the {\it spin-up}. The observed and theoretically expected age-metallicity correlations imply that this increase reflects a rapid formation of the Milky Way disk over $\approx 1-2$ Gyrs.  The transformation of the stellar kinematics as a function of [Fe/H] is accompanied by a qualitative change in chemical abundances: the scatter drops sharply once the Galaxy builds up a disk during later epochs corresponding to [Fe/H]$>-0.9$. Results of galaxy formation models presented in this and other recent studies strongly indicate that the trends observed in the Milky Way reflect {\it generic} processes during the early evolution of progenitors of MW-sized galaxies: a period of chaotic pre-disk evolution, when gas is accreted along cold narrow filaments and when stars are born in irregular configurations, and subsequent rapid disk formation. The latter signals formation of a stable hot gaseous halo around the MW progenitor, which changes the mode of gas accretion and allows development of coherently rotating disk.
\end{abstract}


\section{Introduction}

The goal of Galactic archaeology is to unravel the physics of the high-redshift Universe by studying the close-by surviving relics of the early structure formation. The chain of events connecting the Cosmic Dawn era to the present day Milky Way can be investigated using low-mass stars with lifetimes of order of the Hubble time and which thus bear witness to the first bouts of the early assembly of the Galaxy. The archaeological record is reconstructed by linking stellar elemental abundances and orbital properties whilst grouping stars by their chemical fingerprints, in the approach known as {\it chemical tagging} \citep[see][]{Freeman2002}. 

\citet{ELS} were the first to show how, with the help of galaxy evolution models, observed chemistry and kinematics of nearby stars can be interpreted to peer into the Milky Way's distant past. In their dataset, a tight correlation appeared between stellar metallicity and orbital eccentricity. The metal-poor stars looked like they were on more radial orbits compared to the metal-rich ones which generally adhered to circular motion. However in a follow-up work, using a different observational sample, \citet{Norris1985} pointed out that instead of a simple direct relationship between the orbital circularity and iron abundance, a less clear-cut picture likely existed. They revealed that metal-rich stars could show significant non-circular motions while metal-poor stars were allowed to follow the disk's rotation. Several decades later, the $e$-[Fe/H] plane has been mapped extensively with the help of wide-area sky surveys and shown to be filled with multiple overlapping populations whose exact nature remained unclear until recently \citep[see e.g.][]{Chiba2000,Carollo2010}. 

Thanks to the arrival of the {\it Gaia} \citep[][]{Prusti2016} astrometric data together with large spectroscopic samples, some of the pieces of the Galactic puzzle have been deciphered. It was  shown for example that the Milky Way's stellar halo contains contributions from multiple distinct sources. What had previously been considered a mixed halo "field" can now be differentiated into: some material stripped from small satellites \citep[][]{Myeong_Action,Myeong_velocity,Myeong_Shards,Koppelman2018,Koppelman2019,Yuan2020,Naidu2020,Malhan2022}, copious amounts of tidal debris from a single massive dwarf galaxy \citep[sometimes referred to as Gaia-Sausage or Gaia Enceladus, GS/E; see][]{Belokurov2018,Haywood2018,Helmi2018,Deason2018,Mackereth2019}, as well as many stars kicked out from the disk \citep[][]{Bonaca2017,Gallart2019,Dimatteo2019,Belokurov2020,Amarante2020}. 

Parallel to the charting of the Milky Way's stellar halo - generally represented by stars with low or negative angular momentum - 
significant focus has been on understanding the chemo-kinematic behaviour of the Galactic disk(s). In particular, the enigmatic extension of the so-called thick disk \citep[][]{Gilmore1983} to low metallicities, dubbed the metal weak thick disk (MWTD), has been perplexing the community since the studies of \citet{Norris1985}. While the follow-up analysis by \citet{Morrison1990} appeared to  confirm the earlier claims of \citet{Norris1985} as to the extension of the thick disk into the $-2<$[Fe/H]$<-1$ regime, questions were raised subsequently about possible biases in the photometric metallicities used in both studies \citep[][]{Ryan1995,Twarog1994}. 

In this century, large spectroscopic samples, in particularly those procured by wide-area surveys have provided ample data to study the metal-poor wing of the thick disk. Taking advantage of vast numbers of freshly available spectroscopic metallicities, several independent studies reported an unambiguous detection of the thick disk stars at low iron abundances \citep[see][]{Chiba2000,Beers2002,Carollo2010,Kordopatis2013,Li2017}. Note however that typically in these pre-{\it Gaia} works, the MWTD is always identified as an additional contribution on top of the assumed halo component dominating the stellar number counts at low metallicities. Most recently, several groups used the {\it Gaia} astrometry together with spectroscopic or photometric metallicities to extend the detection of the disk population into very metal-poor regime \citep[][]{Sestito2019,Dimatteo2020,Fernandez2021}. Surprisingly, stars with kinematics similar to both thick and thin disk  are identified, reaching metallicities as low as $-6<$[Fe/H]$<-3$.

There are two observational obstacles to figuring out the exact lineage and the precise timeline of the Milky Way's evolution. First is the absence of high quality age measurements for truly old stars. Asteroseismological data to date remains scarce \citep[see,  e.g.,][]{Montablan2021, Grunblatt2021,Puls2022}, while the resolution of spectro-photometric ages deteriorates rapidly, increasing to 2-3 Gyr beyond 10 Gyr \citep[see e.g.][]{Hidalgo2011, SandersDas}. Second, even the very crude chemical tagging of stars into the "accreted" and the "born in-situ" categories starts to falter at old  ages if only limited chemical information is available. This is simply because outside of the MW bulge, most old stars are metal-poor and would even have comparable $\alpha$ abundances because the $\alpha$ plateau is reached quickly at the onset of star formation. However, as we discuss below in Section~\ref{sec:insitu_selection}, a much more robust identification of the in-situ component is possible using detailed chemical abundance measurements of the APOGEE survey, which serves as a foundation for the present study. 

Great strides in bringing observational picture of the Milky Way assembly into focus have been intertwined with significant improvements in understanding formation of disk galaxies in general \citep[e.g.,][]{naab_ostriker17} and disk of the Milky Way specifically \citep[e.g.,][]{binney13}. Much of the latter theoretical effort was dedicated to investigation of scenarios of the thick disk origin: from the first scenarios involving violent and turbulent early state of the Galaxy naturally leading to a fast formation of a hot disk component \citep{Jones1983, Burkert1992,brook_etal04,bird_etal13,bird_etal21} to formation of the thick disk by depositing stars \citep{abadi_etal03} and/or dynamical heating during mergers \citep[see][]{Quinn1993, Velazquez1999,font_etal01,Hayashi2006, kazantzidis_etal08, Villalobos2008}. 

At the same time, cosmological simulations of galaxy formation revealed that disk formation and evolution constitute an important, but relatively late stage of Milky Way's assembly. Two physically distinct regimes of gas accretion onto galaxies were identified in theoretical models: the ``cold flow'' accretion mode, when gas accretes along warm filaments that can penetrate to the inner regions of halos and a cooling flow mode wherein gas cools onto galaxy from a hot halo \citep{keres_etal05,keres_etal09,dekel_birnboim06,dekel_etal09}.  

During early stages of evolution, when gas accretion is highly chaotic, progenitors of Milky Way-sized galaxies are expected to accrete gas via supersonic narrow filamentary flows. 
The chaotic accretion  leads to irregular, highly turbulent, extended distribution of star forming gas and stars during early stages of star formation in Milky Way-sized progenitors \citep{rosdahl_blaizot12,stewart_etal13,meng_etal19}. It was also realized that feedback in simulations must be sufficiently strong to suppress star formation during these early epochs and prevent formation of massive stellar spheroid  \citep[e.g.,][]{guedes_etal11,hopkins_etal11,hopkins_etal12,stinson_etal13,agertz_etal13,vogelsberger_etal13,crain_etal15}. Strong stellar outflows and bursty star formation during early stages of evolution contribute to the overall chaos in the distribution of dense gas and star formation sites. 

When halo mass and virial temperature become sufficiently large galaxies are able to sustain hot gaseous halo \citep{birnboim_dekel03}. Transition from the early chaotic stage of gas accretion to the slower accretion via cooling of gas from the hot halo or accretion is accompanied by ``virialization'' of gas in the inner regions of halo and formation of coherent long-lived and coherently rotating gaseous disk via a cooling flow \citep{dekel_etal20,stern_etal21}. It is during this stage that the thin stellar disks in spiral galaxies form \citep[][]{hafen_etal22}. 

In this study we aim to probe the early chaotic stages of evolution of the main Milky Way's progenitor and transition to the build-up of the coherently rotating stellar disk. To this end, we use the APOGEE DR 17 data set (Section~\ref{sec:data}) and the available element abundances to identify a sample of stars that likely formed {\it in-situ} in the main progenitor (Section~\ref{sec:insitu_selection}).
We show that stellar kinematics and scatter in element abundances as a function of metallicity reveal a new low-metallicity {\it in-situ} stellar component that we associate with the early turbulent stages of Milky Way's evolution (Section~\ref{sec:ymw}). We also show the signature of the ``spin-up'' of the Milky Way disk in the intermediate metallicity populations. We discuss and interpret observational results in Section~\ref{sec:model_interpretation} and summarize our results and conclusions in Section~\ref{sec:conclusions}. 

\begin{figure*}
  \centering
  \includegraphics[width=0.99\textwidth]{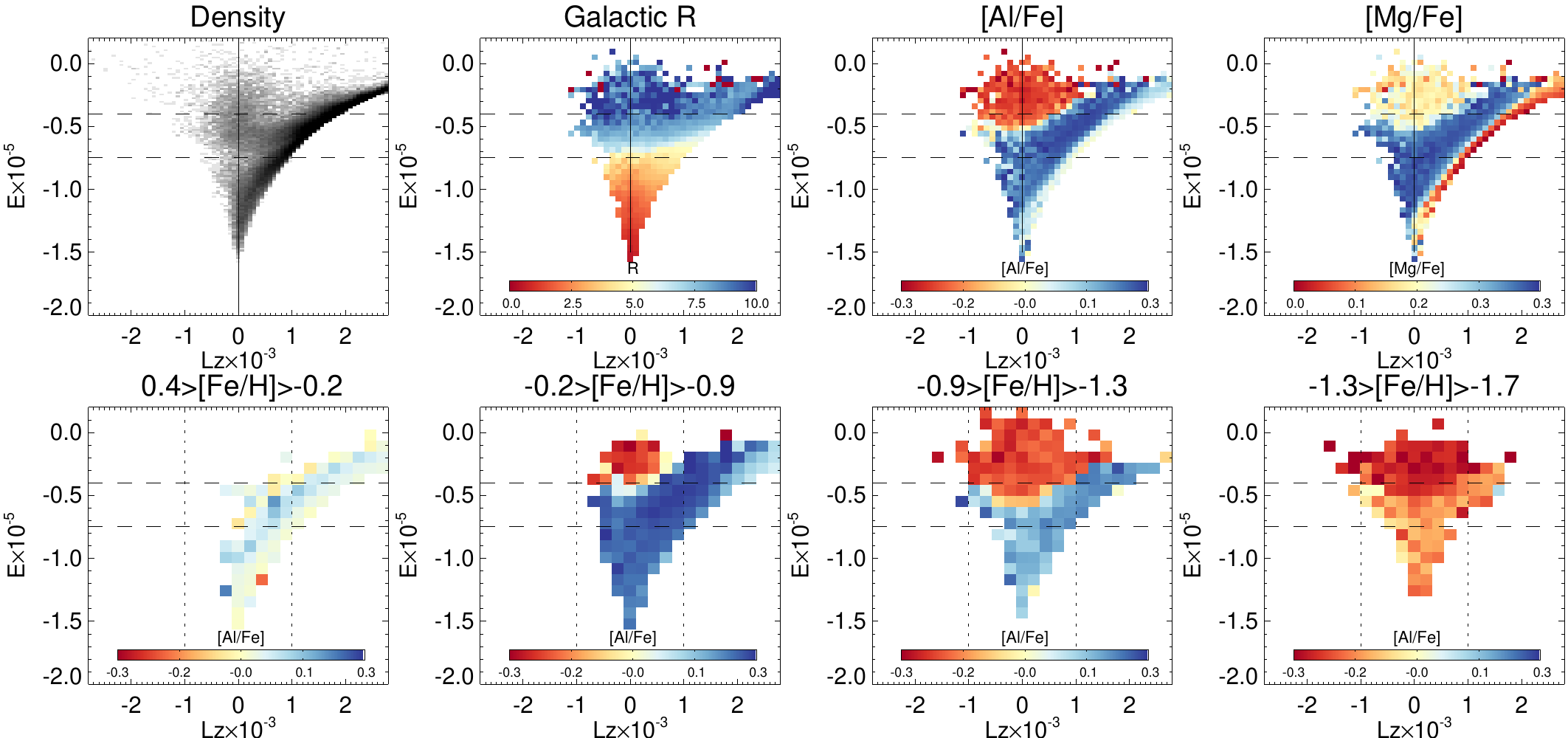}
  \caption[]{Distribution of ADR17 stars in the space of energy $E$ and vertical component of angular momentum $L_{\rm z}$. {\it Top row, 1st panel:} Logarithm of the stellar density. Note a noticeable horizontal gap at around $E\approx -0.75\times10^5$ corresponding the details of the ADR17 footprint. The two dashed horizontal lines show the energy restriction we apply for the subsequent analysis. {\it Top, 2nd:} Median Galacto-centric radius. Note that the stars below the energy gap are located in the bulge region of the Milky Way, i.e. have $R<3$ kpc. {\it Top, 3rd:} Median [Al/Fe]. Note a clear separation of the in-situ (high [Al/Fe] and accreted (low [Al/Fe]) populations. {\it Top. 4th:} Median [Mg/Fe]. The separation between in-situ and accreted is visible by the dynamic range is somewhat reduced. {\it Bottom row} $E, L_{\rm z}$ distribution coloured by [Al/Fe] in bins of iron abundance [Fe/H]. {\it Bottom, 1st:} Only the rotating disk is visible with intermediate [A/Fe]. {\it Bottom, 2nd:} Both accreted and in-situ populations are discernible thanks to very different [Al/Fe]. The in-situ stars display a low-$L_{\rm z}$ wing, known as Splash. {\it Bottom, 3rd:} While the typical [Al/Fe] for the accreted stars is the same at this metallicity, the mean [Al/Fe] of the in-situ population has decreased. {\it Bottom, 4th:} Even though the [Al/Fe] of the in-situ is reduced further, the accreted population can be easily identified since it has an even lower abundance of Al. Note that the shape of the in-situ distribution is now much more symmetric in $L_{\rm z}$ without a prominent rotating component.}
   \label{fig:elz}
\end{figure*}

\section{Milky Way in-situ stars in APOGEE DR17}
\label{sec:data}

In what follows we use aluminium to iron abundance ratios to differentiate between the in-situ and the accreted Milky Way populations following the ideas of \citet{Hawkins2015}. Two aspects of [Al/Fe] evolution combine to help isolate the stars born in a massive galaxy with a rapid chemical enrichment. First is the metallicity dependence of the SN II Al yields \citep[see][]{Woosley1995,Nomoto2013}. Second, similar to $\alpha$-elements, is the dilution of [Al/Fe] with the increased output of SN Ia . 

During early stages of evolution of the Galaxy when metallicity is small the Al output increases with increasing metallicity as  more neutrons become available. While the comparatively massive Milky Way progenitor recycles the polluted gas efficiently and self-enriches quickly, the same process happens at a much more leisurely pace in a dwarf galaxy. The Milky Way enjoys a long enough phase of prolific [Al/Fe] growth before the SNe Ia start to contribute significantly. As a more relaxed dwarf reaches the necessary metallicity levels and starts to churn out Al, SNe Ia kick off and curtail the growth of the [Al/Fe] abundance ratio. Therefore, in the analysis presented below, we view the behaviour of [Al/Fe] at $-2<$[Fe/H]$<-1$ as a basic but powerful indicator of rapid chemical enrichment. As such it is complementary to the classification based on $\alpha$-elements used at higher metallicity \citep[see e.g.][]{Nissen2010,Hawkins2015,Hayes2018}. 

The broad-brush picture painted above has support in the literature. For example, \citet{Hasselquist2021} demonstrated that in the five most massive dwarf galaxies accreted by the Milky Way, namely the Large and Small Magellanic Clouds (LMC and SMC), the Sagittarius, Fornax and the GS/E progenitor, median [Al/Fe] ratios do not exceed [Al/Fe]$\approx-0.2$, while the Galactic disk stars exhibit significantly higher ratios of [Al/Fe]$>0$. The difference in the Al enrichment between the Milky Way and smaller dwarfs was exploited to identify and study the likely accreted halo stars \citep[e.g.][]{Hawkins2015,Das2020,Horta2021}. In contrast, in this study we use it to identify stars formed {\it in-situ} in the main progenitor of the Milky Way. 


\subsection{ADR17 Data}
\label{sec:data_detail}

\begin{figure*}
  \centering
  \includegraphics[width=0.99\textwidth]{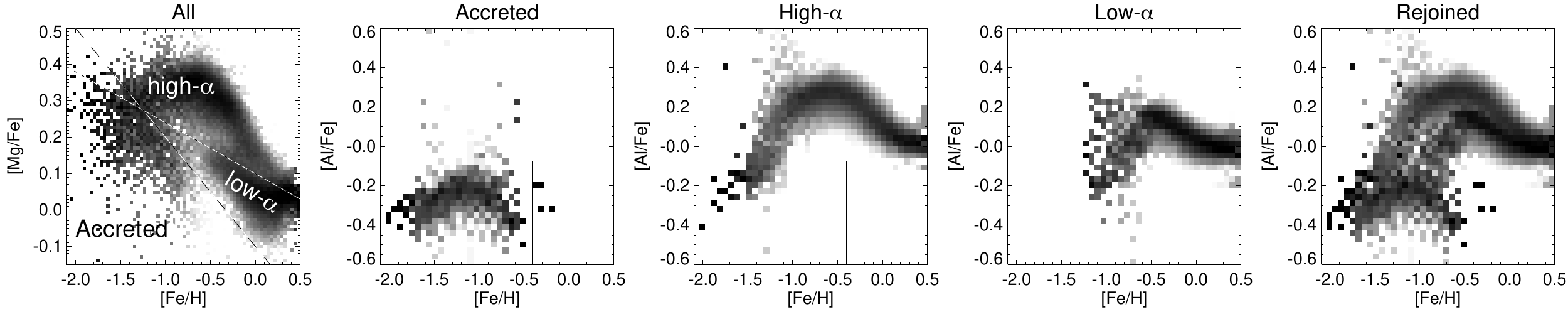}
  \caption[]{{\it First panel:} Column-normalised density of stars in the plane of [Mg/Fe] and [Fe/H]. Dashed and dotted diagonal lines show the selections for the "accreted", "high-$\alpha$" and "low-$\alpha$" populations. {\it 2nd:} Column-normalised density of stars in the "accreted" sample in the [Al/Fe] vs [Fe/H] plane. This sample is dominated by the GS tidal debris. The sequence shows a slow rise up to, followed by a decline and is in agreement with the GS pattern shown in \citet{Hasselquist2021}. {\it 3rd:} Same but for high-$\alpha$ sample. {\it 4th:} Same but for the low-$\alpha$ sample. {\it 5th:} All three sequences added back together.}
   \label{fig:alfe}
\end{figure*}

We rely on the data from the APOGEE Data Release 17 \citep[ADR17,][]{apogeedr17}, more precisely the chemical abundance data is taken from the \texttt{allStarLite} catalogue available on the survey's website. The stellar Galacto-centric coordinates, velocities and integrals of motion as well as their uncertainties are taken from the AstroNN value-added catalogue \citep[see][]{Leung2019,Mackereth2018} which assumes the \texttt{MWPotential2014} gravitational potential from \citet{Bovy2015}. We choose to restrict the analysis to red giant stars with surface gravity estimates of $\log g<3$ and velocity uncertainties of $<50~{\rm km~s}^{-1}$ in each of the three cylindrical velocity components. As we are interested in subtle chemical abundance trends, we also require uncertainty of $<0.2$ dex in the abundance estimates of the 9 elements, C, N, O, Mg, Al, Si, Mn, Fe, Ni, which ADR17 measures most reliably. We also eliminate stars with measurements suspected to be problematic, as signalled by the following flags:
\verb|STAR_BAD, TEFF_BAD, LOGG_BAD|, \verb|VERY_BRIGHT_NEIGHBOR|, \verb|LOW_SNR, PERSIST_HIGH|, \verb|PERSIST_JUMP_POS|,  \verb|PERSIST_JUMP_NEG|, \verb|SUSPECT_RV_COMBINATION|. Duplicate observations are removed by culling all objects with the fourth bit of \verb|EXTRATARG| flag set to 1. Finally, we remove stars within 1 degree of any known Galactic satellite be it a dwarf galaxy or a globular cluster. We specifically remove stars in the Magellanic Clouds by eliminating all objects with \texttt{PROGRAMNAME=magclouds}. The totality of the above selection cuts leaves $\approx$214,000 stars.

Figure~\ref{fig:elz} shows the distribution of the selected stars in the space of energy (E) and Angular Momentum (AM). The bulk of the stars in the sample pile up along the line of the maximal allowed AM. Also visible in the distribution of ADR17 stars is a horizontal gap around $E\times10^{-5}\approx-0.75$ (see the leftmost panel in the top row of the Figure). As already pointed out by e.g. \citet[][]{Lane2021}, this gap is due to the ADR17 footprint on the sky. There are crudely two distinct groups of pointings, one towards the Galactic centre and one sufficiently far away from it. The APDR17 selection function produces a bi-modal energy distribution. This is confirmed in the second panel of the top row where the colour-coding represents the median Galacto-centric distance. As illustrated here, stars with energies $E\times10^{-5}<-0.75$ typically reside within $\sim3$ kpc from the Galactic centre. For most of the analysis presented below we eliminate the "bulge" contribution by removing stars with low energies (as indicated by the lower dashed horizontal line).

Panel 3 (4) in the top row of Figure~\ref{fig:elz} shows the plane of energy and $L_{\rm z}$, the vertical component of AM, colour-coded according to the median abundance of aluminium (magnesium) relative to iron. Both elements exhibit very clear separation between the stars formed in-situ and those formed in smaller progenitor galaxies and subsequently accreted onto the Milky Way. The accreted population typically shows a lower abundance ratio of [Al/Fe] and [Mg/Fe], although the magnesium-to-iron ratio also decreases in the thick disc due to increasingly large contribution of type Ia supernovae; [Mg/Fe] reaches the lowest values for the young stars in the thin disk at the highest metallciities probed.  While both elemental ratios posses some discerning power, aluminium provides a larger dynamic range, in particular at low metallicities. 

The ability of the [Al/Fe] to tell apart the accreted and the in-situ stars has been noticed before \citep[see][]{Hawkins2015} and is illustrated in the bottom row of Figure~\ref{fig:elz}. Here, as above, the $E-L_{\rm z}$ plane colour-coded according to the median ratio of [Al/Fe] in each pixel but for four distinct bins of iron abundance, starting from super-Solar on the left and going to $\approx 0.02 Z_\odot$ on the right. Across the entire metallicity range considered, the in-situ stars are easily differentiated: they have higher [Al/Fe] ratio compared to the accreted population. Even for the stars at the lowest [Fe/H] considered, as shown in the fourth panel, this remains true. Indeed, the in-situ stars here can be seen as a funnel shape structure of lighter colour, corresponding to [Al/Fe]$>-0.1$, while the accreted population forms a higher energy pile mostly of maroon colour, with [Al/Fe]$<-0.1$.

\subsection{In-situ selection}
\label{sec:insitu_selection}

We take advantage of the discriminating ability of [Al/Fe] introduced above to build a sample of stars formed in the Milky Way across a wide range of metallicities. Figure~\ref{fig:alfe} helps to elucidate the details of the [Al/Fe] behaviour of the individual components of the Galaxy as seen by ADR17 in the Solar neighborhood. The Figure connects familiar sequences in the [Mg/Fe]-[Fe/H] plane to their counterparts in the  [Al/Fe]-[Fe/H] space. In the first panel, dashed and dotted diagonal lines are used to separate stars belonging to the accreted, the high-$\alpha$ and the low-$\alpha$ populations. These are then plotted in the subsequent panels. 

The local accreted stars predominantly belong to a single merger event, namely the GES \citep[see e.g.][]{Belokurov2018, Iorio2019}. The accreted [Al/Fe] sequence (panel 2) shows a moderate rise up to $-0.2<$[Al/Fe]$<-0.1$ at around [Fe/H]$\approx-1.2$ followed by a decline. This pattern is entirely consistent with that presented in \citet{Hasselquist2021}. The Milky Way's high-$\alpha$ sequence (panel 3) rises much more rapidly with metallicity and as a result resides mostly above the accreted sequence across the entire range of [Fe/H]. Note that at [Fe/H]$<-1.3$ the high-$\alpha$ sequence dips below [Al/Fe]$\approx-0.1$ and so at lower metallicities the in-situ stars start to overlap with the accreted population. The low-$\alpha$ sequence behaves in a similar fashion albeit offset in metallicity and [Al/Fe]: its rise is less steep and it does not reach as high levels of [Al/Fe] compared to the high-$\alpha$ sequence. After the turn-over around [Fe/H]$\approx-0.5$, the two sequences are almost parallel meeting at super-Solar iron abundances. As the Figure demonstrates, above [Al/Fe]$\approx-0.1$, the contamination of the in-situ sample by the accreted debris is minimal.

\begin{figure*}
  \centering
  \includegraphics[width=0.99\textwidth]{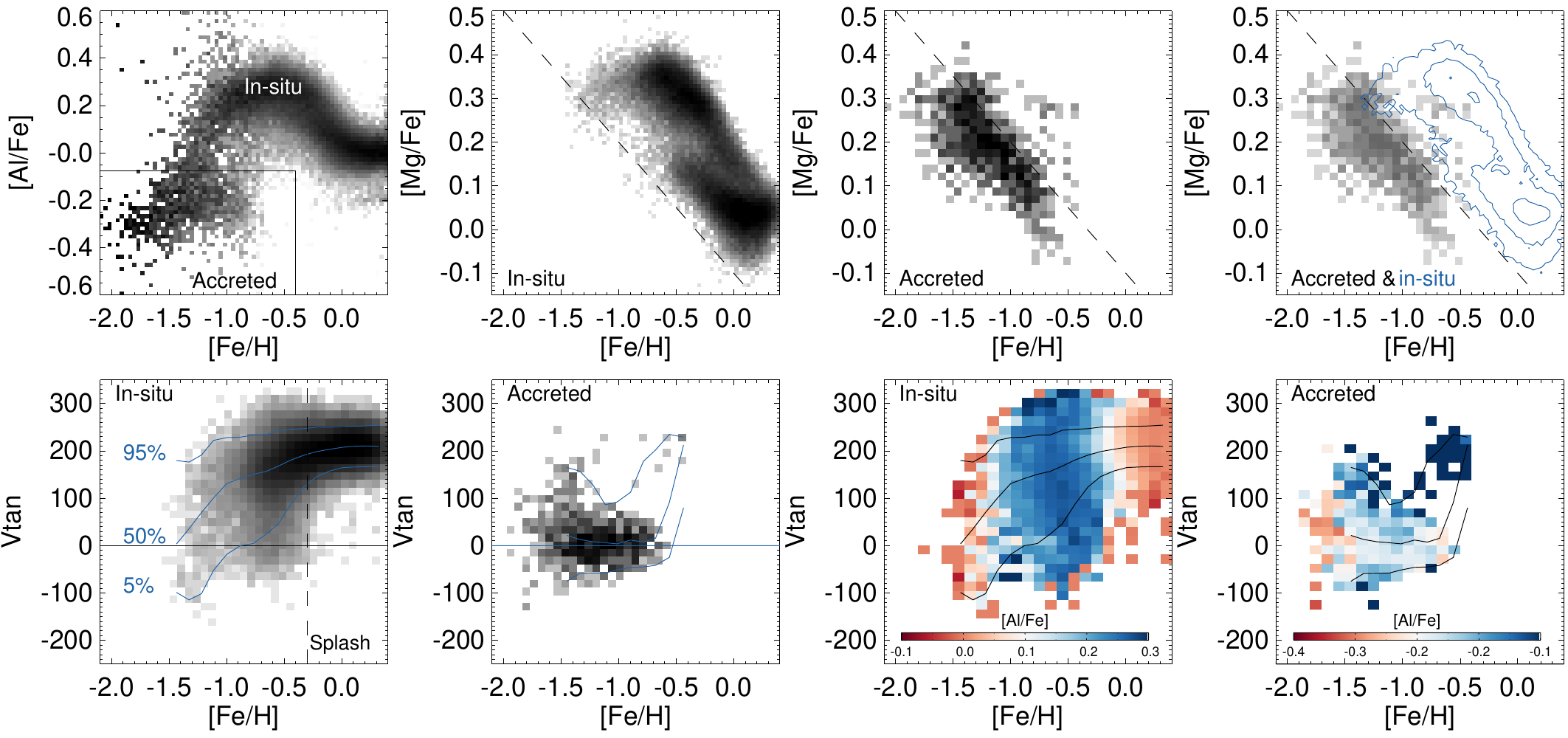}
  \caption[]{Separation of the in-situ and accreted populations. {\it Top row, 1st panel:} Column-normalised density of stars selected as described in Section~\ref{sec:data_detail} with the addition of the energy cuts illustrated in Figure~\ref{fig:elz} in the space of [Al/Fe] and [Fe/H]. Solid black lines show the selection boundaries used to separate the stars born in-situ from those accreted onto the MW. {\it Top, 2nd:} Density distribution of the Al-selected in-situ population in the space of [Mg/Fe] and [Fe/H]. {\it Top, 3rd:} Same for the sample of accreted stars. {\it Top, 4th:} Both in-situ (contours) and accreted (greyscale density) populations are shown. Dashed diagonal line shows an addition cut used to clean the accreted population further. {\it Bottom row, 1st panel:} Logarithm of the density of stars in the plane spanned by azimuthal velocity $V_{\rm tan}$ and metallicity [Fe/H]. Solid curves gives the behaviour of the 5th, 50th and 95th percentiles of the $V_{\rm tan}$ distribution as a function of [Fe/H]. {\it Bottom, 2nd:} Same but for thew accreted stars. Note the difference in the behaviour of the $V_{\rm tan}$ distribution: at low [Fe/H] the in-situ distribution is broader with a rapidly increasing mode. The $V_{\rm tan}$ distribution of the accreted stars is narrow with approximately constant mode close to 0 km~s$^{-1}$. {\it Bottom, 3rd:} The in-situ distribution colour-coded according to the median [Al/Fe] in each pixel. While a clear gradient of [Al/Fe] is seen with [Fe/H], at fixed metallicity no correlation is observed between $V_{\rm tan}$ and [Al/Fe], especially at [Fe/H]$<-1$. {\it Bottom, 4th:} Same for the accreted stars. Note that very high values of $|V_{\rm tan}|$ correspond to an increase of [Al/Fe]. We interpret this as a sign of contamination: some of the in-situ stars with low [Al/Fe] entered the accteted sample.}
   \label{fig:selection}
\end{figure*}

Figure~\ref{fig:selection} illustrates the details of the in-situ selection. To the selection cuts described above we add a restriction on the total energy (as given by the two horizontal dashed lines shown in Figure~\ref{fig:elz}, i.e. $-0.75<E\times10^5<-0.4$) and a requirement that the heliocentric distance does not exceed 15 kpc. This reduces the sample to $\approx$79,500 stars.
The leftmost panel 1 in the top row shows the column-normalised density of the selected stars in the plane of [Al/Fe] and [Fe/H]. Guided by the behavior of local stellar populations shown in Figure~\ref{fig:alfe}, two black lines demarcate the regions predominantly populated by the accreted and in-situ stars. All stars with [Fe/H]$>-0.4$ are deemed to be born in the Milky Way. Below this metallicity,  we allocate stars with [Al/Fe]$<-0.075$ to the accreted population, while the in-situ stars have higher aluminium-to-iron ratios. 

As a sanity check, we show the distribution of stars in the two samples of accreted and in-situ stars defined this way in the plane of [Mg/Fe]-[Fe/H] in the panels 2 and 3 of the top row of the Figure. As panel 2 demonstrates,  for $-1.5<$[Fe/H]$<-0.5$ the in-situ population remains on the alpha "plateau", i.e. the increase in the iron abundance is accompanied by a constant Mg/Fe ratio. This can be contrasted with the evolution of magnesium in the sample of accreted stars. This relatively local sample is dominated by the tidal debris from the GES merger. The [Mg/Fe] track of the GES stars is very different from that of the stars born in the Milky Way's main progenitor. In the range of metallicities considered, the GES-dominated halo population evolves to lower values of [Mg/Fe] with increasing iron abundance, starting the decline already at around [Fe/H]$\approx-1.3$. This is a signature of significant SN Ia contribution to the Fe abundance and a symptom of a slower self-enrichment in a dwarf galaxy progenitor whose mass is significantly lower than that of the MW. This pattern is in perfect agreement with the earlier studies of the $\alpha$ abundance in the GS debris \citep[see][]{Helmi2018, Haywood2018,Mackereth2019}. 


The bottom row of Figure~\ref{fig:selection} demonstrates the chemo-kinematic behaviour of the two populations selected [Al/Fe]. Specifically, the bottom row panels show the variation of the cylindrical azimuthal velocity component $V_{\rm tan}$ (equivalent to cylindrical $V_{\rm \phi}$) as a function of metallicity. First (leftmost) panel gives the density of in-situ stars in the plane of $V_{\rm tan}$ and [Fe/H] together with three lines corresponding to the run of the 5th, 50th and 95th percentiles of the velocity distribution. According to the density distribution and confirmed by the behaviour of the velocity percentiles, the Milky Way possessed very little coherent spin at the epoch corresponding to the metallicity of ${\rm [Fe/H]}\lesssim-0.9$. The median azimuthal velocity of the in-situ population is built up from approximately $0<V_{\rm tan}<50$ km~s$^{-1}$ at ${\rm [Fe/H]}\approx-1.5$ to a substantial $\approx150$ km~s$^{-1}$ at ${
\rm [Fe/H]} \approx-0.9$. In contrast, the accreted halo stars (mostly the GES debris) have  $V_{\rm tan} \approx 0$ km~s$^{-1}$ for the entire [Fe/H] range probed (note that here we apply an additional cut in the [Mg/Fe] plane shown in the top row in an attempt to clean the accreted sample of a small number of in-situ interlopers). 

Note that the mode and the width of the distribution of azimuthal velocities of the accreted population is largely due to the GS progenitor. The dwarf's orbit probably experienced a rapid radialization \citep[see an extensive discussion in][]{Vasiliev2021} thus depositing its stellar debris on orbits with a typical AM close to zero. The width of the $V_{\rm tan}$ velocity distribution is relatively narrow, i.e. $50<\sigma_{\rm tan} ({\rm km~s^{-1}})<100$  corresponding to the total mass of the parent GES halo  of $\lesssim 10^{11} M_{\odot}$. Curiously, for the accreted stars with $-1.5<$[Fe/H]$<-1$ an additional component with high $V_{\rm tan}$ is visible reaching azimuthal velocities in excess of 100 km~s$^{-1}$. These peculiar and abrupt changes in the velocity distribution is the results of the contamination of the accreted sample by the low-metallicity in-situ stars. As Figure~\ref{fig:alfe} demonstrates the high-$\alpha$ sequence enters the accreted [Al/Fe] region around [Fe/H]$\approx-1.3$. Although such prograde rotating stars could in principle be a relic of a satellite accreted nearly in the disk plane \cite[e.g.,][]{abadi_etal03}, this is unlikely in this case as the stars with significant $V_{\rm tan}$ have also a relatively high [Al/Fe] ratio as we discuss below.

\begin{figure}
  \centering
  \includegraphics[width=0.49\textwidth]{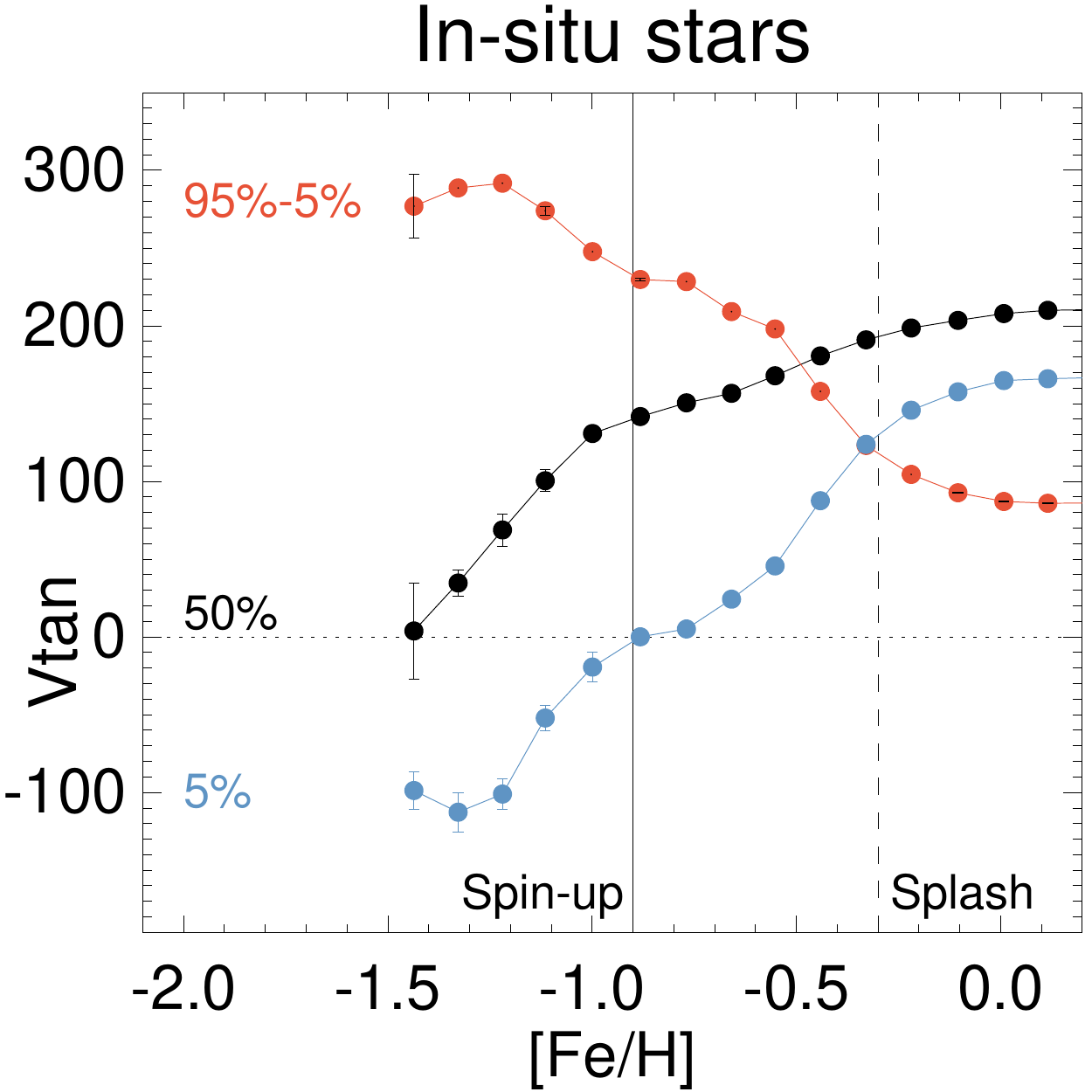}
  \includegraphics[width=0.49\textwidth]{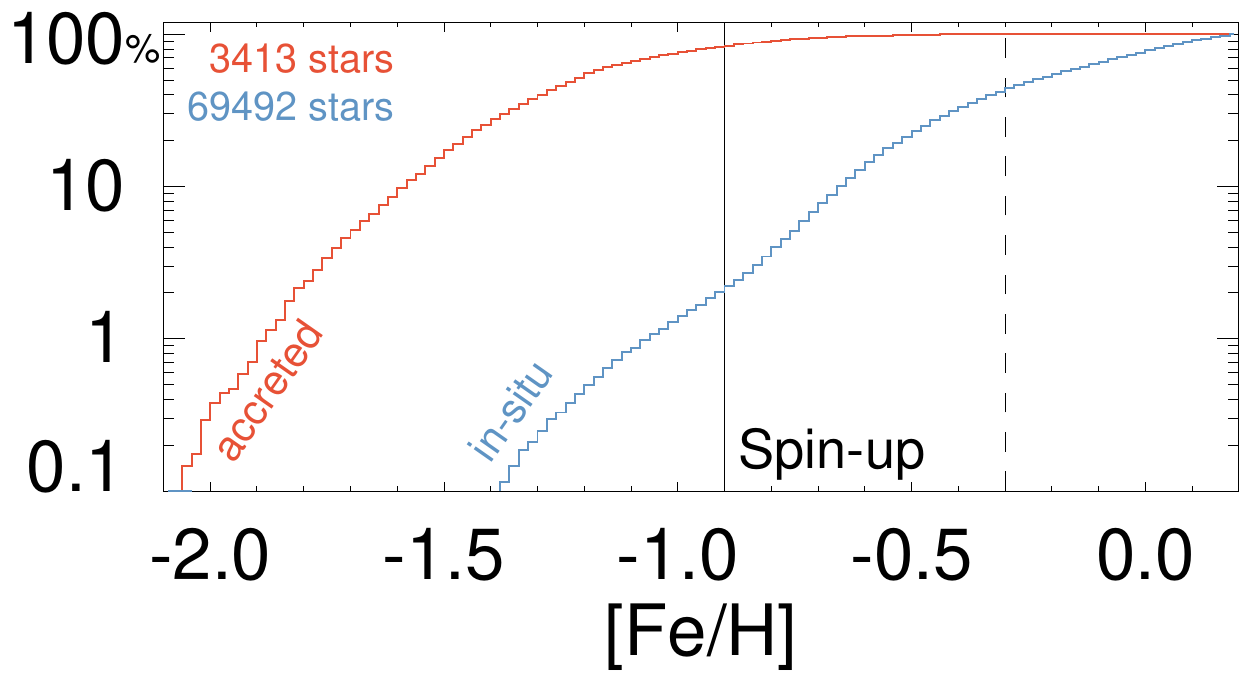}
  \caption[]{{\it Top:} Evolution of the in-situ azimuthal velocity distribution as a function of metallicity. Black curve and filled circles with error-bars give the change of the mode of the distribution, while blue and red showing the low-$V_{\rm tan}$ wing and the $3.3\sigma$ extent correspondingly. Culmination of the Spin-up (Splash) phase is shown with a vertical solid (dashed) line. During the Spin-up the overall spin of the in-situ stars is rapidly increasing, while the Splash is characterised by the emergence of the low-$V_{\rm tan}$ wing of the distribution. The bulk of the disk remains intact and spinning during and after the Splash. {\it Bottom:} cumulative metallicity distribution of the in-situ (accreted) sample is shown as the red (blue) histogram. Note the drastic difference between the two populations. At low [Fe/H], the in-situ sample is significantly depleted compared to the accreted one. This is indicative of a much faster and a more efficient star-formation activity in the MW progenitor at early times. By the time the Spin-up phase was over, MW only formed $\approx1\%$ of its stars. The accreted sample shows a slow-down above [Fe/H]$\approx-1.3$ and truncation at [Fe/H]$\approx-0.5$ corresponding to the epoch of the strong interaction between the GS progenitor and the MW.}
   \label{fig:velocity}
\end{figure}

Signatures of potential cross-contamination between the samples of accreted and in-situ stars can also be explored in panels 3 and 4 of the bottom row of Figure~\ref{fig:selection}. Here, the velocity distribution as a function of metallicity is colour-coded according to the median [Al/Fe]. Panel 3 shows that while there is a global trend of [Al/Fe] as a function of metallicity in the in-situ stars, at fixed metallicity [Al/Fe] ratio does not appear to be correlated with the azimuthal velocity. This is not the case for the accreted stars shown in the panel 4 of the Figure. The regions corresponding to the highest $V_{\rm tan}$ also show a clear increase in the [Al/Fe] (compared to the rest of the population). This is a clear sign of contamination of the accreted sample with a small number of in-situ stars. Their presence is easy to recognise at high positive azimuthal velocities, but it is also betrayed at high negative values of $V_{\rm tan}$ where a mild increase in [Al/Fe] can also be seen. This is likely because in the range of $-1.5<$[Fe/H]$<-0.9$ the in-situ population has a much broader azimuthal velocity dispersion compared to that of the accreted halo. As discussed above the accreted population's azimuthal velocity spread is constrained by the internal kinematics of the GS progenitor. As judged by the patterns revealed in panels 3 and 4 of Figure~\ref{fig:selection}, we conclude that while the accreted population is mildly contaminated at [Fe/H]$<-1.3$, there are no obvious signs of the in-situ sample contamination.

\section{The early Milky Way}
\label{sec:ymw}

\subsection{Kinematics}
\label{sec:kinematics}

Figure~\ref{fig:velocity} shows dependence of the kinematic properties in the Al-selected in-situ population as a function of metallicity, as well as fractions of stars in each sample at metallicity smaller than a given value. The curves corresponding to the 5th and 50th percentiles in the top panel are given to illustrate the change in the lowest and the typical AM population. This is complemented with the curve showing the difference between 95th and 5th percentiles representing the $3.3\sigma$ extent of the $V_{\rm tan}$ distribution. There are evident changes in the behaviour of the azimuthal velocity distribution as all three curves undergo marked transitions at two critical metallicities. Note that behavior shown in the three curves is similar but not identical, which indicates that the alterations the Milky Way's kinematics incurred were non-trivial. 

\begin{figure*}
  \centering
  \includegraphics[width=0.99\textwidth]{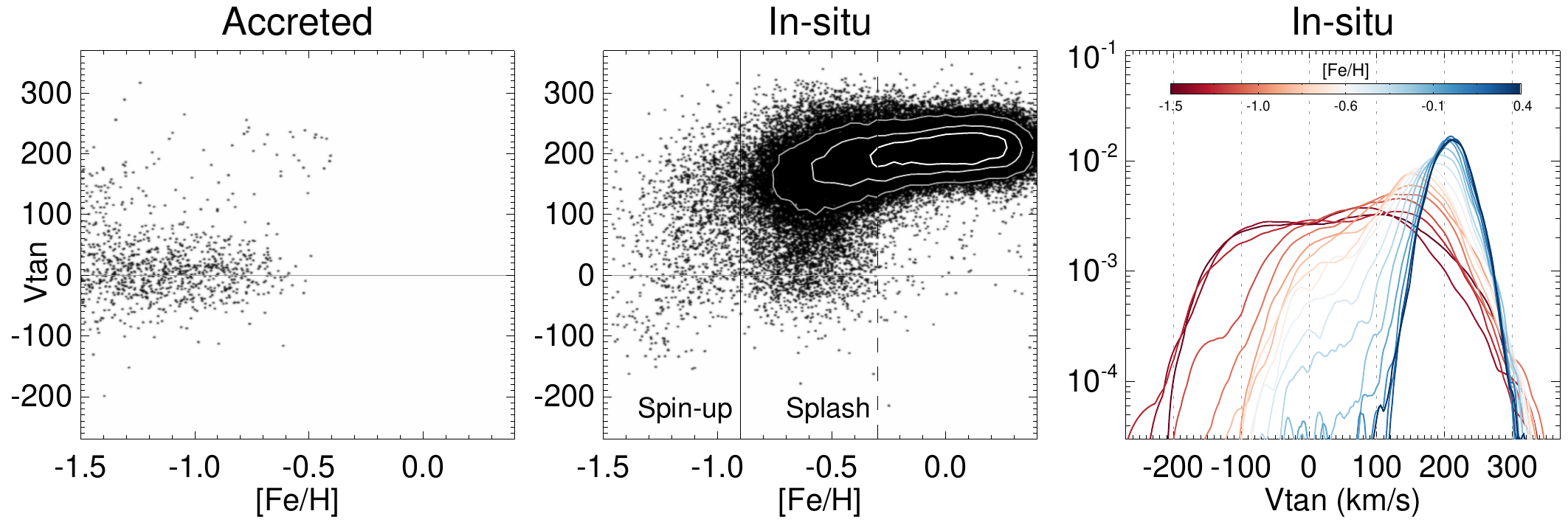}
  \caption[]{Azimuthal velocity $V_{\rm tan}$ as a function of metallicity. {\it Left:} Stars identified as accreted using [Al/Fe]. This sample is dominated by the GS debris with the median spin near zero and a narrow range of $V_{\rm tan}$. Note the clear contamination at [Fe/H]$<-1.3$ from the in-situ stars (see also Figure~\ref{fig:alfe}). {\it Middle:} Scatter plot of the in-situ stars. {\it Right:} KDE (with Epanechnikov kernel of optimal size) estimates of the probability density function of $V_{\rm tan}$ in bins of [Fe/H]. Note the low net spin and the wide span of the in-situ azimuthal velocity distributions at low metallicities.}
   \label{fig:vel_feh}
\end{figure*}

Before we delve into the details of the chemo-kinematic evolution of the Milky Way population, let us briefly compare the cumulative metallicity distributions of the stars born inside and outside the Milky Way. These are given in the bottom panel of the Figure. At metallicities $-2<$[Fe/H]$<-1.5$, the accreted stars dominate by $>2$ orders of magnitude.\footnote{Note that the in-situ sample's metallicity distribution is truncated due to the [Al/Fe] cut applied. This selection bias does not affect the overall differences between the distributions as described here.} This is because at fixed metallicity we are peering further back into the past in the MW progenitor as it formed stars faster and self-enriched more efficiently. The small fraction of in-situ stars at low metallicities is qualitatively consistent with expectations of modern galaxy formation models, as we discuss in Section~\ref{sec:model_interpretation} (see Fig.~\ref{fig:fstar_cum_model}). At [Fe/H]$>-1$, while Milky Way kept producing stars, the SF activity in the GES progenitor (the dominant contributor to the accreted sample) was stifled and quickly shut down as it interacted with the much bigger MW. 

Let us follow the velocity curves from the metallicity just above ${\rm [Fe/H]}=0$, i.e. not too far from the present day, to lower [Fe/H], traversing Figure~\ref{fig:velocity} from right to left. This should approximately correspond to older stellar ages due to observations metallicity-age relation (and model expectations, see Section~\ref{sec:model_interpretation}). Stars with $-0.3<$[Fe/H]$<0.2$ reveal little change in their kinematics: all three curves are flat. The overall spin (black) is at its highest, $V_{\rm tan}>200$ km~s$^{-1}$ (note that this is an average of the azimuthal velocities of the stars at different Gaalacto-centric distances). The total ($3.3\sigma$) extent of the azimuthal velocity distribution is at its lowest, $<100$ km~s$^{-1}$ meaning that the average azimuthal velocity dispersion (red) is $<30$ km~s$^{-1}$. The velocity distribution is tight around the peak: the lowest $V_{\rm tan}$ stars (blue) are only $\sim 30$ km~s$^{-1}$ lower than those in the mode of the distribution. This is the recently formed portions of the Milky Way disk: kinematically cold and spinning fast.

The first dramatic transition happens at [Fe/H]$\approx -0.3$. The low AM wing of the distribution quickly stretches to lower velocities; it decreases by a factor of 2 at [Fe/H]$\sim-0.5$ and finally reaches 0 km~s$^{-1}$ at [Fe/H]$\approx-0.75$. The expansion of the low-$V_{\rm tan}$ side of the distribution is reflected in the growth of the overall spread, which changes from $\sim100$ km~s$^{-1}$ at [Fe/H]$\approx-0.3$ to $\sim200$ km~s$^{-1}$ at [Fe/H]$\approx-0.7$. Note that the creation of a noticeable low-AM wing is not accompanied by a strong decrease in the typical spin of the Galaxy. Over the same range of metallicity, i.e $-0.7<$[Fe/H]$<-0.3$, the median $V_{\rm tan}$ decreases only by $\sim50$ km~s$^{-1}$. The stretching of the velocity distribution towards low and even negative azimuthal velocities at intermediate iron abundances has been noticed before. The lowest AM stars are attributed to the so-called in-situ stellar halo \citep[see e.g.][]{Bonaca2017, Gallart2019}. This structure is hypothesised to have been created by heating and splashing of the existing Milky Way disk as it interacts with the GES progenitor galaxy \citep[see][]{Dimatteo2019, Belokurov2020,Grand2020,Dillamore2021}.

Then after a brief lull at $-0.9<{\rm [Fe/H]}<-0.7$ where all three curves flatten, the second metamorphosis occurs between [Fe/H]$\approx-1.5<$ and [Fe/H]$\approx-0.9$. The most remarkable change is reflected in the behaviour of the Galactic spin: at the lowest of the metallicities considered i.e. [Fe/H]$\approx-1.5$, no significant net rotation is present. This is also the epoch where the most substantial retrograde population is recorded together with the widest spread in azimuthal velocities. The change from no spin to a fast rotation happens rather quickly with increasing metallicity: by [Fe/H]$\approx-0.9$ the Milky Way's typical $V_{\rm tan}$ is already $\approx150$ km~s$^{-1}$. Note that while the changes in the changes in the 5th and the 50th percentiles are also fast and striking, the evolution of the spread of the $V_{\rm tan}$ distribution is milder. As the Galaxy spins up, the spread tightens slowly: it only drops by less 20\%, from $\approx$300 km~s$^{-1}$ to $\approx$250 km~s$^{-1}$ as the metallicity increases from -1.5 to -0.9. Nonetheless, the azimuthal velocity distribution is at its wildest when $-1.5<$[Fe/H]$<-1.3$: the $3.3\sigma$ interval covers at least 300 km s$^{-1}$.

Figure~\ref{fig:vel_feh} provides additional details of the change in the $V_{\rm tan}$ distribution with metallicity. Azimuthal velocities of individual stars are shown as a function of [Fe/H] for the Al-selected in-situ population in the middle panel. At the lowest metallicities probed no obvious disk population can be detected. Note however that given the simple [Al/Fe] cut applied, many of the genuine in-situ stars with [Fe/H]$<-1.3$ will be a part of the accreted sample. This is indeed the case as shown in the left panel of the Figure: at [Fe/H]$<-1.3$ the $V_{\rm tan}$ distribution for the accreted sample suddenly becomes broader, including many stars with $V_{\rm tan}>100$ km s$^{-1}$. The right panel of Figure~\ref{fig:vel_feh} presents $V_{\rm tan}$ probability distribution functions for small bins of [Fe/H]. These pdfs are obtained using kernel density estimation (KDE) with Epanechnikov kernel of optimal size. The pdfs corresponding to the lowest metallicity bins are very broad, spanning $\approx500$ km s$^{-1}$ from 200 to 300 km s$^{-1}$. The Figure emphasises that the Spin-up, i.e. the change from a broad distribution with little net spin to a much narrower peak centered on a $V_{\rm tan}>100$ km s$^{-1}$ happens rapidly. 

The velocity ellipsoid of the in-situ stars with $-1.5<$[Fe/H]$<-1.3$ appears isotropic, i.e. $\sigma_{\rm z}\approx\sigma_{\rm R}\approx\sigma_{\rm \phi}$, with all three dispersions in the range of 80-90 km s$^{-1}$. This is distinct from the 3D velocity distributions of both the GES debris and the Splash. The GES radial velocity dispersion is twice as high, in the range of 170-180 km s$^{-1}$. In the Splash, the azimuthal dispersion is lower, around 50-60 km s$^{-1}$ \citep[][]{Belokurov2020}.

\begin{figure*}
  \centering
  \includegraphics[width=0.99\textwidth]{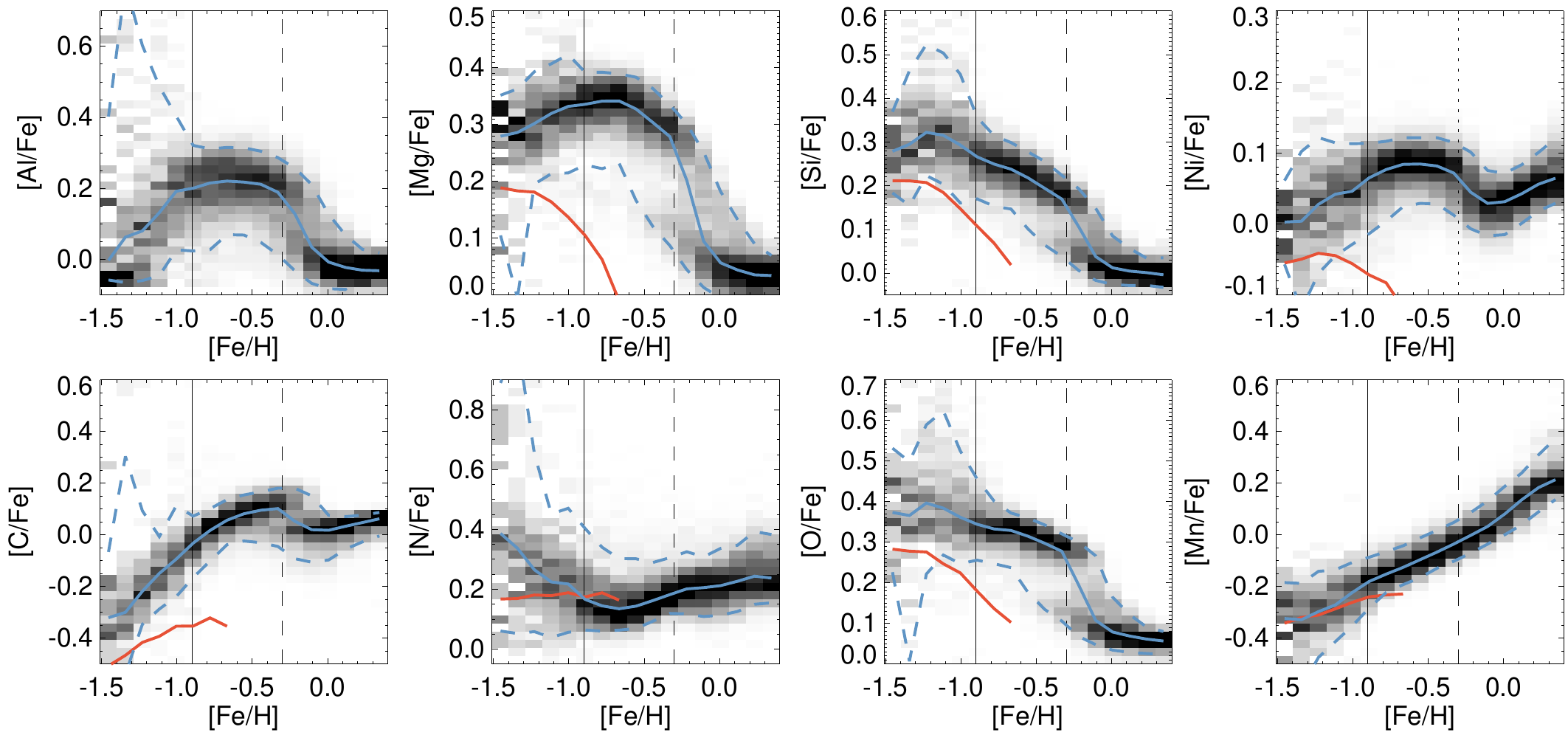}
  \caption[]{Column-normalised density of the distribution of the elemental abundance ratio [X/Fe] of the in-situ stars as a function of metallicitty [Fe/H]. Blue (red) solid curve shows the median of the distribution of the in-situ (accreted) population. Additional cuts in energy $E\times10^5>-0.4$ and tangential velocity $|V_{\rm tan}|<70$ km s$^{-1}$ are applied to the accreted sample to minimse the contamination. Dashed blue curves show 5\% and 95\% of the {\it in-situ} distribution.}
   \label{fig:chem_den}
\end{figure*}

The stars born in the Milky Way before it spins up, i.e. settles into a coherent disk provide a window into the earliest phases of our Galaxy's evolution. We therefore name this population \name, to symbolise the dawn of the Milky Way's formation. The \namesp and the Splash stellar populations correspond to the two junctures in the Milky Way's youth when rapid and consequential changes are unleashed. From the kinematic perspective, these two phases, that of formation and that of disruption, appear unconnected. The biggest give-away is the evolution of the spin: even though the interaction with the GS/E progenitor crumbles and disperses some of the existing stellar disk, the bulk of it remains intact retaining its relatively high rotational velocity. The creation of the Splash from the pre-existing disk also helps to explain the difference in the amount of low- and negative $V_{\rm tan}$ stars during the two epochs. It is plausible to expect that a distinction between the \namesp and the Splash can also be drawn in the chemical dimension: the vastly different kinematic structure of the young Milky Way before and after disk formation may leave an imprint in the behaviour of key elemental abundances. In what follows we take advantage of the exquisite quality of the ADR17 data to examine chemical signatures of the Spin-up and the Splash phases.

\subsection{Chemistry}
\label{sec:chemistry}

In order to make the chemical analysis more robust and to minimise the effects of any unwanted trends with stellar surface gravity and/or effective temperature, in this Section we apply a stricter cut $\log g<2$ to define stellar sample. Note that narrowing the range further to $1<\log g<2$ does not change any of the conclusions drawn below but increases the uncertainty.

\subsubsection{Trends with [Fe/H]}

Because several channels with distinct timescales contribute to the production of iron, using Fe as a reference helps to emphasise temporal evolution of the star-formation and enrichment process. 

Figure~\ref{fig:chem_den} shows the metallicity dependence of the [X/Fe] abundance ratios for the eight chemical elements measured most reliably as part of the ADR17, namely: Al, Mg, Si, Ni, C, N, O and Mn. These greyscale 2D histograms give column-normalised densities for the Al-selected sample of in-situ stars, or in other words, abundance ratio probability distributions conditional on [Fe/H]. In addition, blue solid (dashed) lines show the behaviour of the mode (5th and 95th percentiles) of the distribution. These can be compared to the median trends for the accreted population shown as the red solid curves. 

The element abundance ratios of the in-situ stars display several clear trends. Around the metallicity of the Splash ([Fe/H]$\approx-0.3$) two distinct sequences are visible at a given metallicity in all of the elements considered apart from N and Mn. The clearest separation between the high and the low abundance ratio branches can be seen in [Al/Fe], [Mg/Fe], [Si/Fe] and [O/Fe]. For stars with [Fe/H]$<-0.3$ formed before the Splash,  the upper branch is more populated, while for the stars with [Fe/H]$>-0.3$ the lower branch has the higher density. The accreted stars generally follow different trends in all elements. In all elements but N, the in-situ and the accreted abundance trends start to converge at the lowest metallicities. The [N/Fe] sequences are overlapping at around [Fe/H]$\approx-1$ but are set apart by 0.1-0.2 dex at [Fe/H]$=-1.5$.

\begin{figure*}
  \centering
  \includegraphics[width=0.99\textwidth]{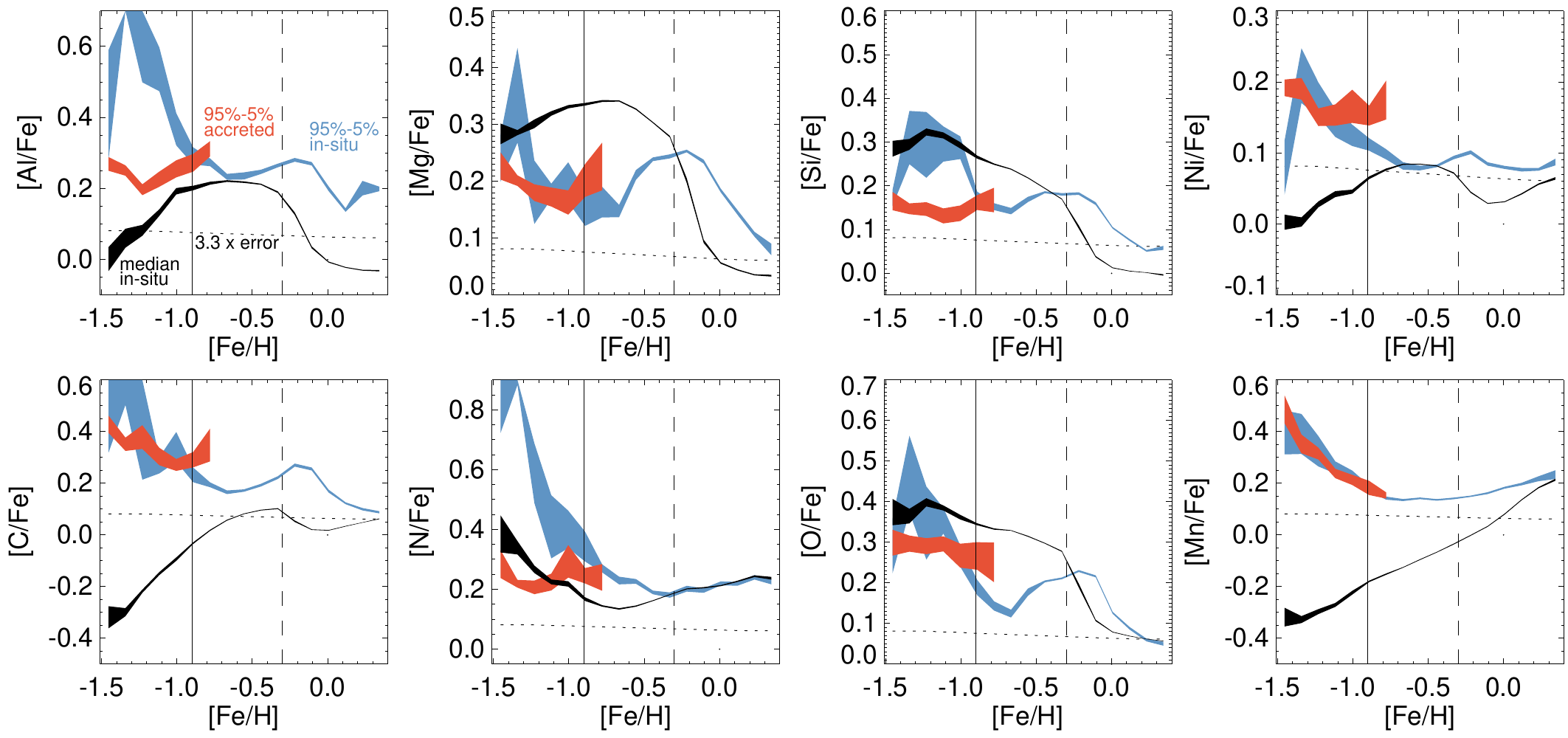}
  \caption[]{Statistics of the abundance ratio [X/Fe] distribution of the in-situ stars as a function of metallicity [Fe/H]. The width of the band shows the associated bootstrapped uncertainties. Black band is the median of the distribution, while the blue band gives the $3.3\sigma$ extent. For comparison, red band gives the abundance spread in the accreted population. The width of the abundance ratio distribution can be compared with the size of the ADR17 measurement uncertainty of the abundance ratio: dotted curve shows 3.3 times the measurement error.}
   \label{fig:chem_stat}
\end{figure*}

A standout feature of the abundance trends shown in Figure~\ref{fig:chem_den} is the abrupt increase of the abundance spread at metallicities below [Fe/H]$\approx-0.9$ (in several elements, the gradual uptick in dispersion happens at slightly higher metallicity, i.e. [Fe/H]$<-0.7$). As the Figure demonstrates, many tight sequences that are consistently well-behaved at higher metallicities appear smeared or even completely destroyed below the threshold which we identified kinematically in the previous Section. The most dramatic increases in the abundance dispersion at [Fe/H]$<-0.9$ are seen in [Si/Fe], [Ni/Fe], [N/Fe] and [O/Fe]. 

To investigate this phenomenon further, Figure~\ref{fig:chem_stat} shows the trends of the abundance distributions as a function of [Fe/H]. Here the median of the [X/Fe] abundance ratio (black) can be compared to the extent of the distribution (blue for in-situ and red for accreted stars, $3.3\sigma$ spread mapped by the difference between the 95th and the 5th percentiles). The behaviour of each element is shown as a band whose width gives the associated uncertainty (estimated using the bootstrap method). For the abundance dispersions we subtract the median of the reported uncertainties in quadrature. Note however that these uncertainties are extremely small (typically $<0.05$ dex across the entire metallicity range probed), as indicated by the black dotted lines that show $4\times$ the median uncertainty for each [Fe/H] bin. 

Figure~\ref{fig:chem_stat} confirms that in all eight elements the spread in the abundance ratio of the in-situ population undergoes a dramatic change at ${\rm [Fe/H]}\approx -0.9$. At the same time, the median abundance track does not show any obvious large scale variations around the same metallicity in all elements but Al. Aluminium is shown to be sharply increasing up to [Fe/H]$=-0.9$ where it plateaus and stays constant until the contribution of type Ia supernovae starts to be noticeable at around [Fe/H]$\approx -0.5$. The behaviour of the abundance spread for the Splash population is completely different: for stars with $-0.7<$[Fe/H]$<-0.3$ no strong abundance dispersion variation can be seen. Some minor spread increase is detectable but this is only because in the this metallicity range, two narrow sequences are present in many of the elements considered. 
Comparing the chemical signatures of the \namesp and the Splash, two distinct patterns are clear. The Splash stars share properties of the disk population, indicating their likely disk origin. At $-0.7<$[Fe/H]$<-0.3$, setting aside the apparent bi-modality of many of the elemental abundance ratios, the chemical enrichment sequences as reported by ADR17 are very narrow. Correspondingly, the $3.3\sigma$ abundance extent of the Splash stars is at the level of 0.1-0.2 dex. The elemental abundance spread in the stars born during the Spin-up phase is a factor of 2-4 larger. The cleanest example is nitrogen. Here no [N/Fe] bi-modality (i.e. existence of two standalone sequences at fixed [Fe/H]) is present at intermediate metallicities thus the abundance spread for stars with [Fe/H]$=-0.9$ is approximately constant at the level of 0.2 dex. The dispersion of [N/Fe] ratio grows sharply at lower metallicities reaching 0.8 dex at [Fe/H]$<-1.3$. Note that the median [N/Fe] also shows a mild increase (of order of 0.2 dex) at low metallicities. 

Based on their distinct chemical behaviour, as characterised by the extent of the abundance distribution, for the stars above and below [Fe/H]$=-0.9$ threshold we argue that the \namesp and the Splash populations are not connected. As we discuss in detail in Section~\ref{sec:model_interpretation}, the low metallicity in-situ stars were not born in a disk and later dynamically heated, but had been born during the earlier Spin-up epoch under conditions sufficiently different from those that existed in a cold and fast-rotating disk later.

The overall extent shown in Figure~\ref{fig:chem_stat} gives an over-simplified illustration of the abundance ratio distribution as it does not indicate changes of its shape. As Figure~\ref{fig:chem_den} demonstrates for many elements (e.g., [N/Fe] or [O/Fe]) the increase in the abundance ratio spread is clearly asymmetric. We explore the asymmetries in the distributions of chemical abundance ratios in the Appendix~\ref{sec:asymmetry}. Figure~\ref{fig:chem_stat_pm} studies the upper and lower branches (with respect to the mode) of each of the [X/Fe] distributions separately. The Figure confirms that at low metallicity, i.e. at [Fe/H]$<-0.9$, distributions of [Si/Fe], [N/Fe] and [O/Fe] become very asymmetric. In [Si/Fe] and [N/Fe] the bulk of the spread is upwards from the mode, while for [O/Fe] the behaviour of the two sides of the distribution appears to switch around [Fe/H]$\approx-1.3$. 

\begin{figure}
  \centering
  \includegraphics[width=0.49\textwidth]{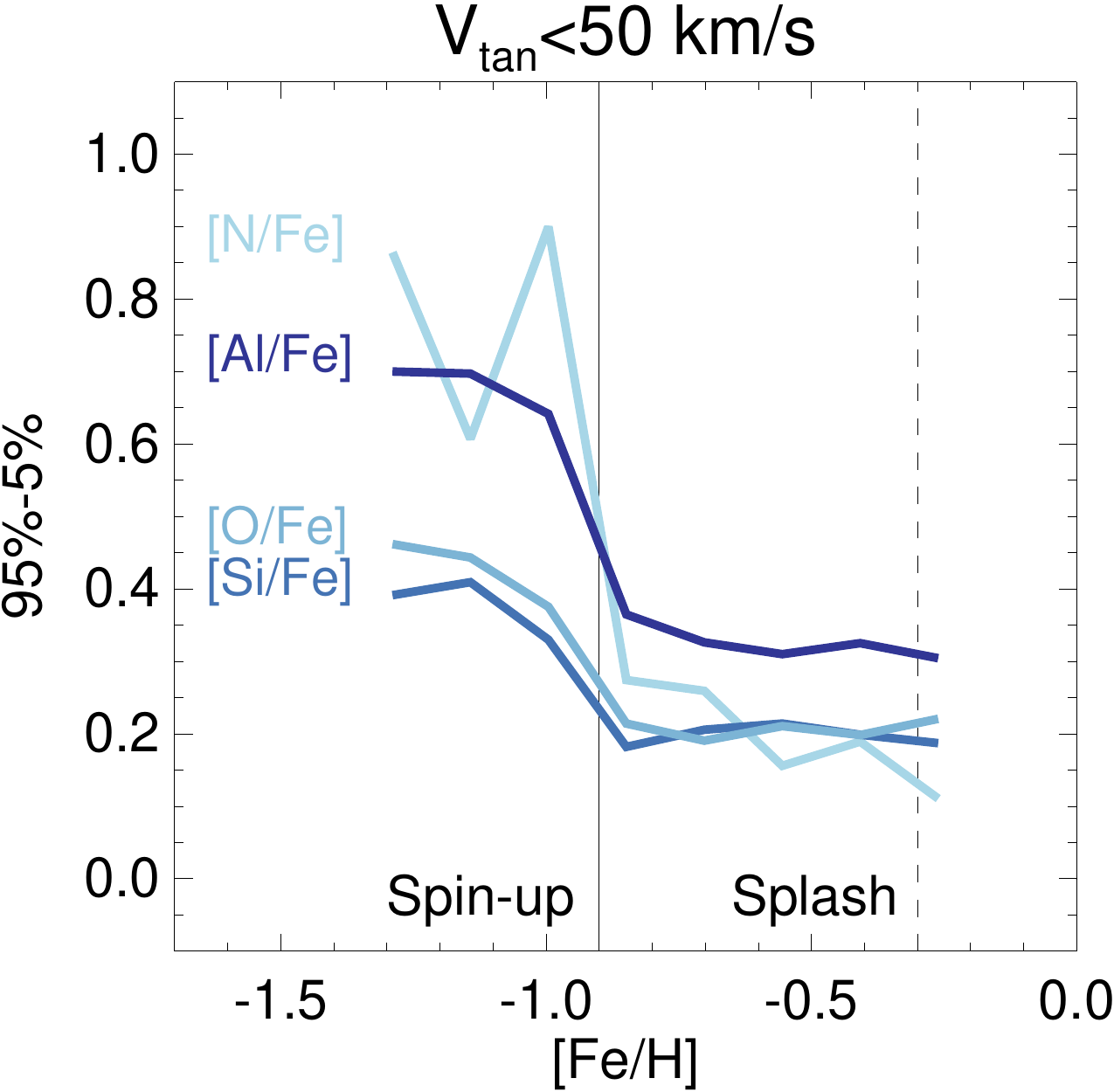}
  \caption[]{Evolution of the spread of the abundance ratio distribution of [Al/Fe], [Si/Fe], [N/Fe] and [O/Fe] as a function of metallicity for the in-situ stars with low or negative angular momentum, i.e. $V_{\rm tan}<50$ km s$^{-1}$. At high metallicity this sample is dominated by the Splash population.}
   \label{fig:chem_retro}
\end{figure}

At high metallicity, the distribution of the chemical abundance ratios of the in-situ population  shown in Figure~\ref{fig:chem_stat} is dominated by the Galactic disc stars. This makes it difficult to assess whether the Splash population behaves differently from \namesp i.e. that born before (and during) the Spin-up phase. To clarify this, Figure~\ref{fig:chem_retro} displays the change in the abundance dispersion (as estimated by the 95\%-5\% difference) for the in-situ sample limited to retrograde stars only, i.e. to stars with $V_{\rm tan}<50$ km s$^{-1}$ \citep[c.f.][]{Belokurov2020}. Here only the four elements that showed the increase in dispersion compared to the accreted populations are shown, namely Al, Si, N and O. Notwithstanding the reduced number of stars, the picture remains the same. In fact, for the retrograde-only stars the behaviour is simplified: the metal-rich portion with [Fe/H]$>-0.7$ shows no bumps or wiggles, staying flat, while the increase in the width of the distribution at ${\rm [Fe/H]}<-0.9$ is sharper.

\subsubsection{Trends with [Mg/H]}

Unlike iron, the Mg production is dominated by the core-collapse supernovae (CCSN). Thus, as pointed out by \citet{Weinberg2019}, using Mg as a reference element greatly simplifies chemical enrichment trends \citep[see also][]{Wheeler1989,Cayrel2004,Hasselquist2021}. For the elements mostly produced by CCSN, such as oxygen,  abundance ratio relative to Mg stays constant if the yields are not metallicity-dependent. For odd-Z elements with metallicity-dependant yields, such as aluminum, the [X/Mg] ratio will steadily increase with [Mg/H]. For iron-peak elements, where multiple sources can contribute to the production (e.g., Fe), trends with [Mg/H] will be straighter compared to the trends with [Fe/H]. Overall, across all elements, trends of the abundance ratio [X/Mg] with [Mg/H] will look simpler, flatter and less dependent on the details of the star-formation history. 

\begin{figure}
  \centering
  \includegraphics[width=0.49\textwidth]{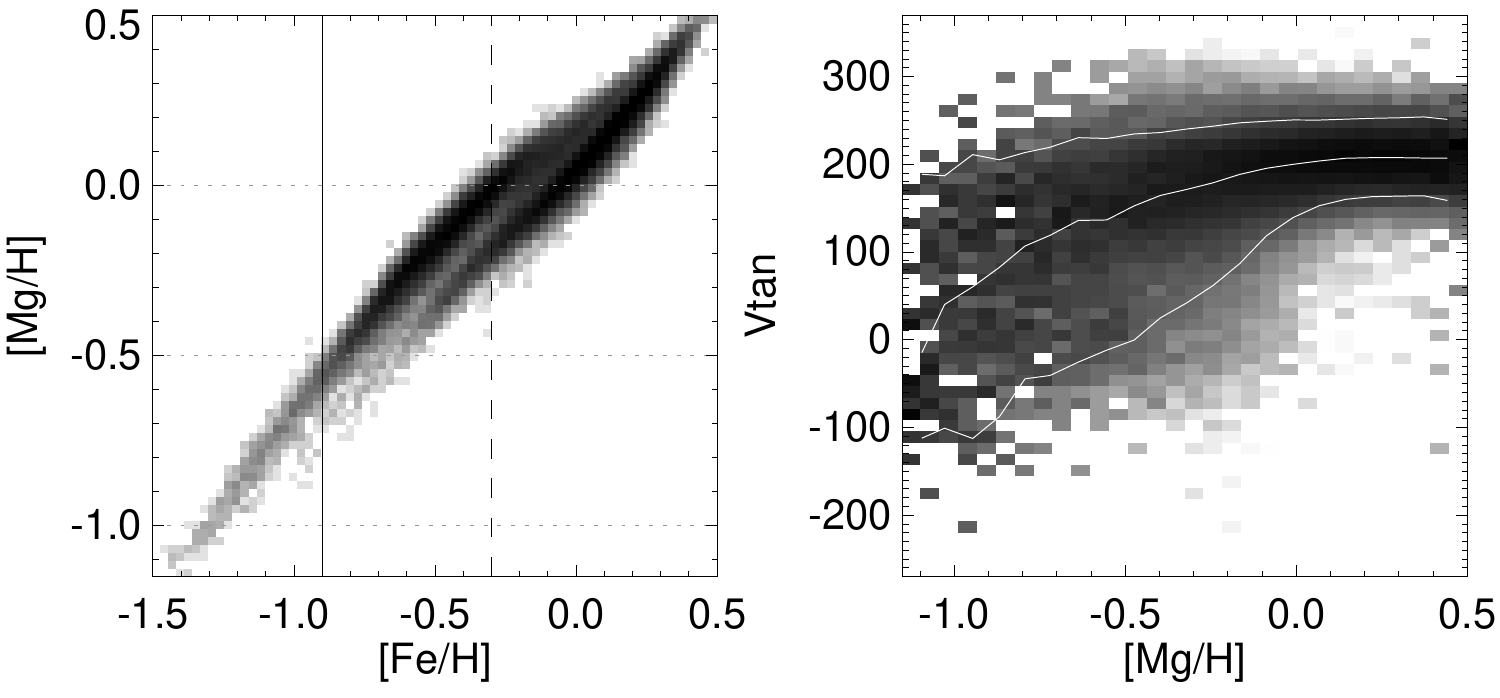}
  \caption[]{{\it Left:} Magnesium abundance as a function of iron abundance. Logarithm of stellar density in the space of [Mg/H] and [Fe/H] is shown for the in-situ sample. Across a wide range of metallicity, at fixed [Mg/H], two values of iron abundance are possible thanks to the contribution of two iron sources, CCSNe and SNe Ia. {it Right:} Logarithm of column-normalised density in the space of azimuthal velocity and [Mg/H] for the in-situ sample. White lines mark 5\%, 50\% and 95\% of the $V_{\rm tan}$ distribution. At low [Mg/H] the transition for low $V_{\rm tan}$ to high $V_{\rm tan}$ is less sharp due to the increased contribution of the disk component at low [Mg/H].}
   \label{fig:mgh_feh}
\end{figure}

To help relate the [X/Mg] abundance trends to the [X/Fe] ones discussed above, Figure~\ref{fig:mgh_feh} shows the relation between [Mg/H] and [Fe/H]. Note that the relation is not one-to-one thanks to enrichment periods with multiple Fe sources in action. As a result, the threshold metallicity of [Fe/H]$=-0.9$ corresponds to a range of [Mg/H] values, namely $-0.7<$[Mg/H]$<-0.5$. This leads to a blurring of the Spin-up kinematic transition discussed above, as illustrated in the right panel of Figure~\ref{fig:mgh_feh}.

\begin{figure*}
  \centering
  \includegraphics[width=0.99\textwidth]{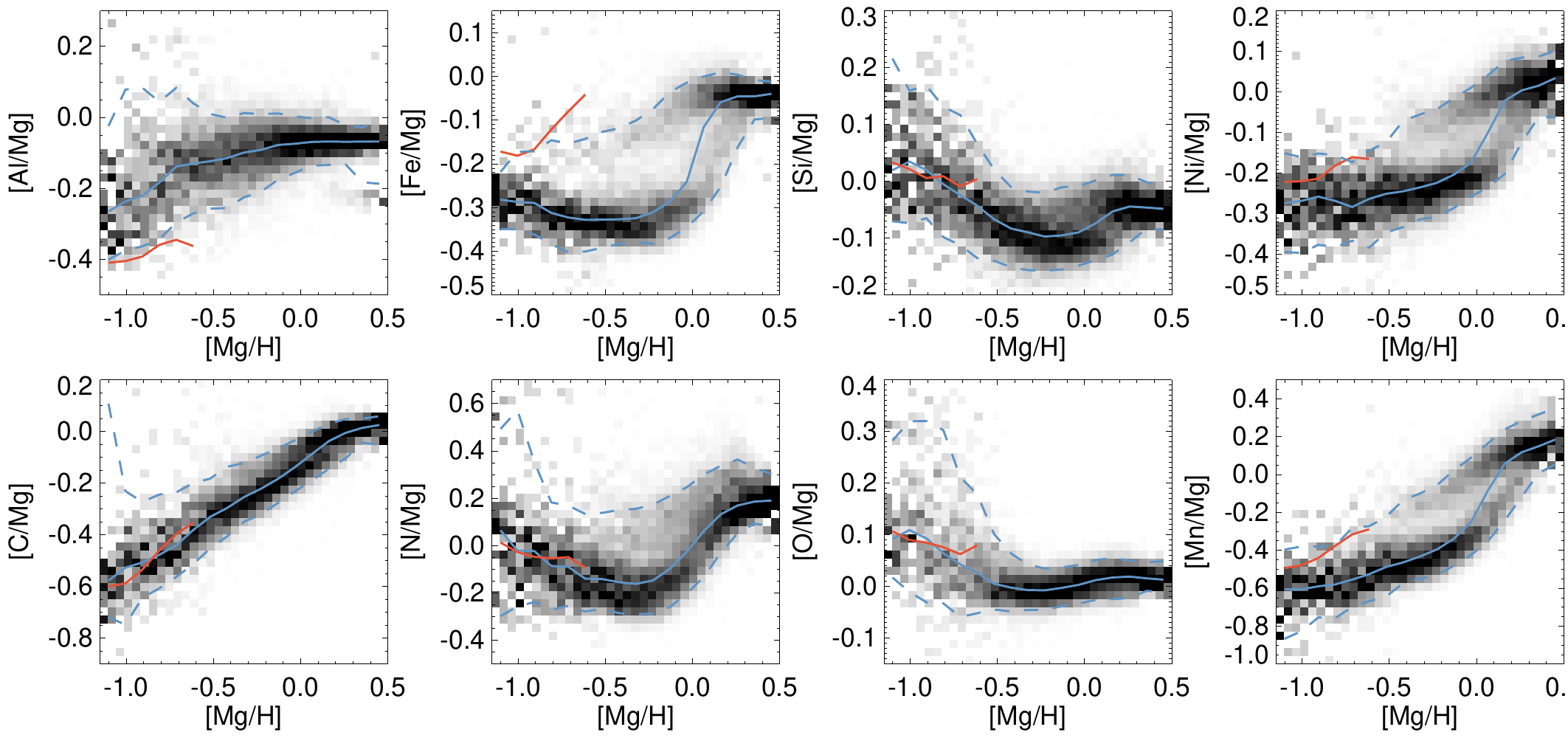}
  \caption[]{Same as Figure~\ref{fig:chem_den} but for the evolution of the abundance ratios referenced to Mg. Note a flatter appearance of most tracks and a better match between the in-situ and the accreted populations.}
   \label{fig:xmg_density}
\end{figure*}

Figure~\ref{fig:xmg_density} shows the column-normalised density of stars in the space of [X/Mg] vs [Mg/H] similar to Figure~\ref{fig:chem_den}. In line with the ideas discussed in \citet{Weinberg2019}, most of the elemental abundance trends are significantly flatter, with the exception of C and N that show a larger amplitude of variation. Moreover, the  abundance tracks of the in-situ and the accreted populations are much more alike, with the median curves lying close to (or on top of) each other. This includes Si, Ni, N and O that previously showed unambiguously dissimilar behaviour. For Fe, Ni and Mn, the accreted and the in-situ abundance tracks are offset  thanks to a higher contribution of SN Ia at fixed metallicity in a dwarf galaxy environment compared to that in the Milky Way. Finally, the accreted population is some 0.1 dex lower in [Al/Mg] compared to the in-situ. We have checked that this is not an artefact of the selection (as the two populations are split on [Al/Fe]) and the difference persists if the accreted stars are selected using orbital inforamtion only. If Al was produced by CCSN only similar to Mg, we would expect similar levels of [Al/Mg] in the accreted and in-situ stars. Note that both Al and Mg are also produced by AGBs and in small quantities in SN Ia. Looking at the behaviour of other diagnostic elements dominated by either of these two sources, e.g. Ni for SN Ia and N for AGB, it is not clear that either could produce such an offset in [Al/Mg] between the accreted and in-situ populations.

\begin{figure*}
  \centering
  \includegraphics[width=0.99\textwidth]{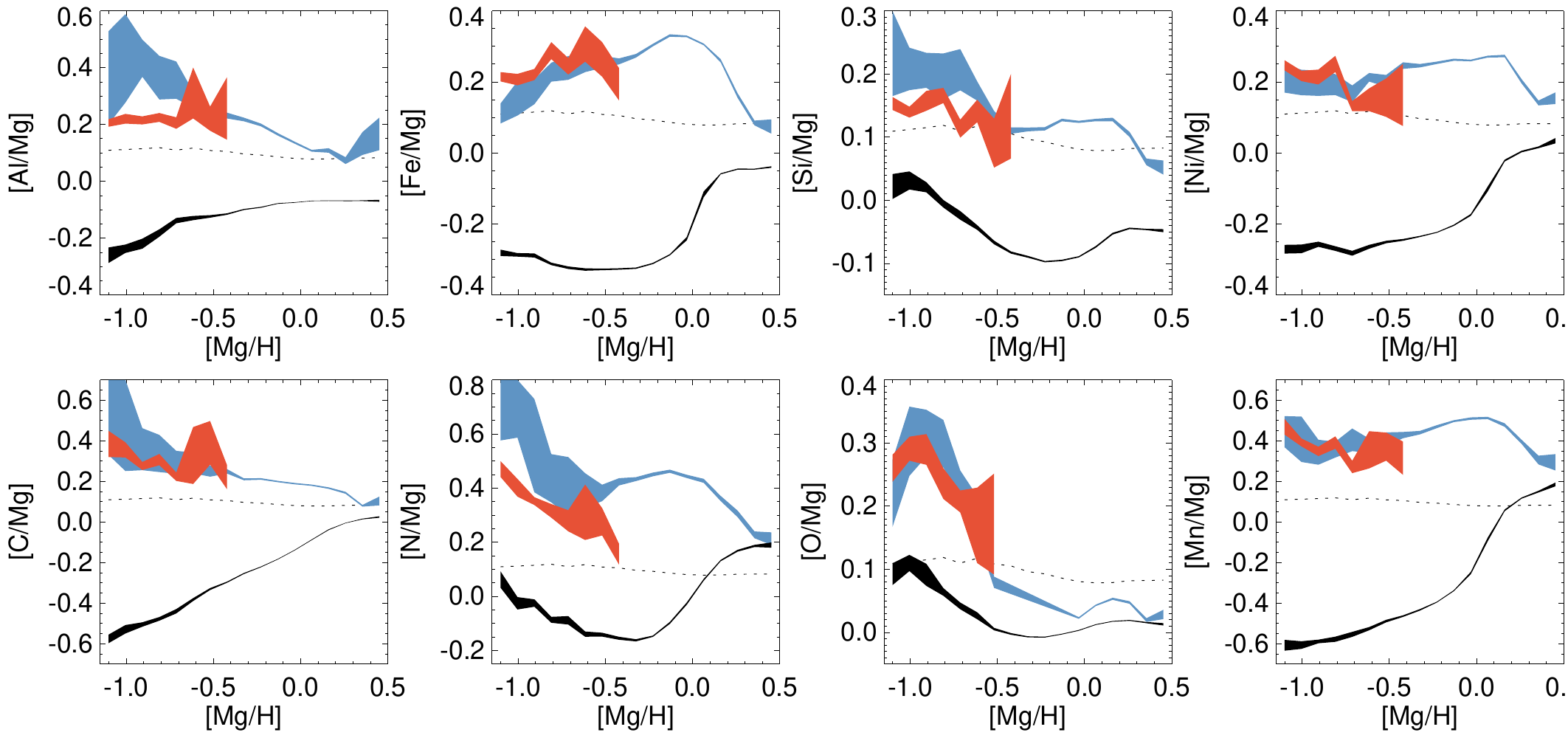}
  \caption[]{Same as Figure~\ref{fig:chem_stat} but for the evolution of the distribution of the abundance ratios referenced to Mg. Note the excess dispersion in [Al/Mg], [Si/Mg] and [N/Mg] at low [Mg/H].}
   \label{fig:xmg_stats}
\end{figure*}

As Figure~\ref{fig:xmg_stats} demonstrates, not only the median tracks are flatter for [X/Mg] ratios, the spread increase towards lower metallicity is greatly reduced for the in-situ stars across all elements. For the accreted population, the abundance dispersion remains flat for all elements but N and O, likely due to the varying contribution of AGB stars. The relative reduction of the abundance spread when referenced to Mg makes sense: [X/Mg] ratios track the relative build up of metals while [X/Fe] fluctuates with rapid changes in gas accretion, star formation rate, and gas depletion time \citep[e.g.,][]{johnson_weinberg20}, as discussed in more detail in Section~\ref{sec:chem_scatter_models}. 

Comparison of the trends of scatter in the elemental abundance ratios referenced to Fe and Mg indicates that much of the increase in scatter at low metallicities is due to stochasticity of [Fe/H] due to the burstiness of the early star-formation histories of both the MW and the dwarf progenitor of the accreted population (see discussion in Section~\ref{sec:model_interpretation}). Despite the simpler behaviour of [X/Mg] ratio trends, the increase of scatter in  [Al/Mg], [Si/Mg] and [N/Mg] of the in-situ stars is still present at metallicities below the metallicities corresponding to the Spin-up stage of $-0.7<$[Mg/H]$<-0.5$. 


\subsubsection{Similarities with chemical anomalies in star clusters}
\label{sec:gc_anomalies}

In the in-situ sample, the largest persistent increases of scatter of the element ratios at low metallicities are those displayed by Al and N. This behaviour is reminiscent of large chemical anomalies observed in stars in most of the Galactic globular clusters (GCs). Although very uniform in [Fe/H], GCs have turned out to have significant spreads in He, C, N, O, Na and Al, with other elements such as Mg and Si also implicated, which indicate multiple star formation episodes during cluster formation \citep[see][for review]{Gratton2004,bastian_lardo18}. Enrichment in some elements is facilitated by the consumption of others, e.g. N and Na are produced while C and O are depleted, leading to high-amplitude abundance (anti)correlations within a single GC. In some GCs, overproduction of Al, sometime by more than 1 dex, has been linked to the depletion of Mg, thus establishing strong Al-Mg anti-correlation \citep[see,  e.g.,][]{Pancino2017,Masseron2019}. 

It is hypothesised that the necessary enrichment takes place in the interiors of the earlier generations (often referred to as the first generation or 1G) of the GC stars and the synthesised material subsequently pollutes the older generations (referred to as the second generation or 2G). As of today no clarity exists as to how and where exactly the synthesis happens \citep[][]{Renzini2015,bastian_lardo18}. Some of the requirements may be very unusual: for example the transformation of Mg into Al necessitates extreme temperatures achieved in massive main sequence stars, in excess of 100 $M_{\odot}$ \citep[see e.g.][]{Prantzos2007, Prantzos2017, Denissenkov2014} or potentially, AGBs \citep[see][]{Cottrell1981}. It is also unclear how exactly the products enriched by the 1G stars are delivered to their 2G siblings. 

\begin{figure*}
  \centering
  \includegraphics[width=0.99\textwidth]{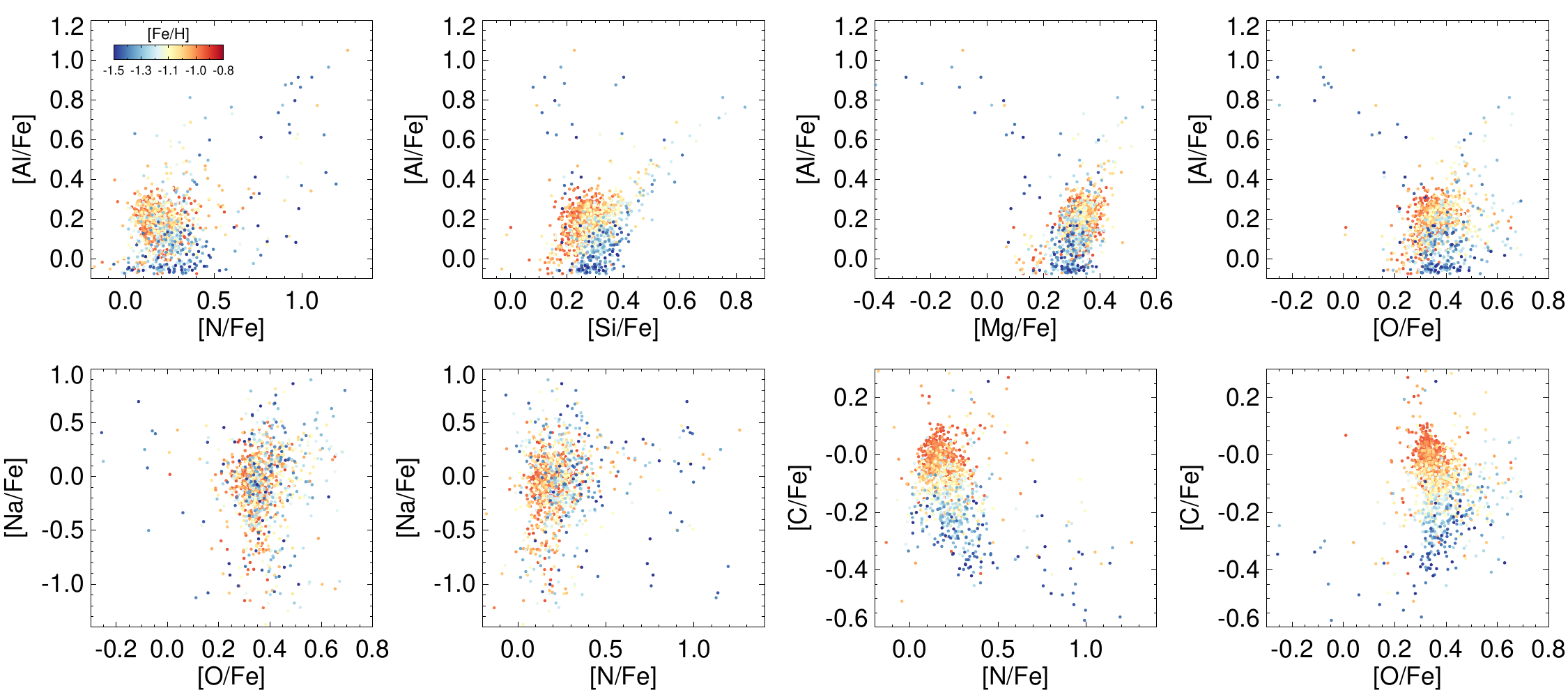}
  \caption[]{Abundance ratio correlations for the in-situ stars with $-1.5$[Fe/H]$<-0.8$. The data-points are colour-coded according to their metallicity. At low metallicity, three groups with distinct Al behaviour are noticeable. First, a small group of high [Al/Fe] outliers, showing correlation with [N/F] and anti-correlation with [Mg/Fe] and [O/Fe]. Second, slightly larger group with [Al/Fe] values lower than the first but still in excess of [Al/Fe]$>0.4$. These stars show a correlation between [Al/Fe] and [Si/Fe]. Finally the third group has [Al/Fe]$<0.4$ and exhibiting correlations between C and N, as well as between C and O. The N dispersion of this group is reduced compared to the high Al, N outliers but is still substantial.}
   \label{fig:al_si_n}
\end{figure*}

Moreover, instead of a well-defined trend, GCs exhibit a range of behaviours: for example, sometimes a correlation between is Mg and Al is observed, sometimes it is not \citep[e.g.][]{Carretta2009, Pancino2017}; in some clusters Al correlates with Si and in others it does not \citep[e.g.,][]{Yong2005, Carretta2009, Masseron2019}. However, it has been noted that the amplitude of the spread and the strength of the abundance correlation grows with the increasing cluster mass and decreasing metallicity \citep[e.g.,][]{Carretta2010, Masseron2019}. The metallicity dependence is particularly striking, instead of a smooth trend a discontinuity exists around [Fe/H]$\approx-1$ below which the overal [Al/Fe] spread increases sharply \citep[see, e.g.,][]{Pancino2017}.

Although no fully developed theory exists to explain the genesis of the chemical anomalies observed in the Galactic GCs, signatures of N and Al over-abundance have been used to identify stars that may have belonged to massive star clusters in the past but are now in the field. In some of the most convincing implementations of the "chemical tagging" idea, excess of N and Al (and sometimes Si) have all been used to single out former GC member stars in the field \citep[see e.g.][]{Martell2011, Martell2016, Lind2015,Schiavon2017, Koch2019, FT2020}.

Do the in-situ stars contributing to the excess Al, N, Si abundance spread at [Fe/H]$<-0.9$, i.e. in the \namesp population and during the Spin-up phase discussed earlier in this paper behave similarly to the stars born in the Galactic GCs? Figure~\ref{fig:al_si_n} provides an answer to this question by exploring possible correlations in abundance ratios of the MW stars with $-1.5<$[Fe/H]$<-0.8$. Focusing on the behaviour of the Al abundance shown in the top row, at least three groups of stars with distinct properties can be identified. First, there are the clear outliers -- the stars with [Al/Fe]$>0.4$ that exhibit a correlation with [N/Fe] and an anti-correlation with [Mg/Fe] and [O/Fe]. These stars, mostly limited to [Fe/H]$<-1.3$, display exactly the chemical signatures observed in the MW GCs. Extreme in some of the chemical properties this group is rather small, counting a dozen or so objects. A second slightly bigger group with a somewhat higher metallicity and with lower values of [Al/Fe] on average shows a different behavior: no clear correlation between Al and N, as well as the absence of anti-correlation between Al and Mg, O. This group shows a positive correlation between Al and Si, with an excess of Si$>0.4$. 
Finally, the third group comprising the bulk of the low metallicity stars has lower values of [Si/Fe]$<0.4$. This group shows weaker but still discernible correlations between Al and Si, C and N, O and C. 

The first two groups contribute the most to the increase of the abundance dispersion at [Fe/H]$<-0.9$ in Al and Si. In particular, the first group with the chemical trends most in agreement with the observed GC anomalies contains the highest Al (as well as Mg and O) outliers. Note that even if these stars are removed, the dispersions in [Al/Fe] and [Si/Fe] remain anomalous at [Fe/H]$<-0.9$. However removing the stars from both the first and the second group removes the increase of scatter in [Al/Fe] and [Si/Fe] entirely. The increase of scatter in [N/Fe], however, persists as it is driven by the stars from the third (most numerous) group, with relatively low N abundances (as demonstrated by the top left panel of Figure~\ref{fig:al_si_n}. The bottom row of the figure also shows the behaviour of Na as a function of O and N. Unfortunately, no clear pattern can be detected, due to the relatively low quality of Na measurements in ADR17 at this level of [Fe/H]. The last two panels show the behaviour of C, N and O, and confirm the patterns identified above.

To conclude, only a small fraction of stars, namely those in the first group discussed above (largest Al and N abundances), show unambiguously the behaviour consistent with the GC origin. Note however that the fractional contribution of these stars increases sharply  below the metallicity [Fe/H]$\approx-1$ and drives the excess dispersion in [Al/Fe]. This sharp metallicity dependence is very similar to the pattern displayed by anomalous stars in the Galactic GCs \citep[e.g.,][]{Pancino2017} and in the field \citep[see, e.g.,][]{Schiavon2017}. Moreover, given the diversity of the chemical behaviour observed in the Galactic GCs we can not rule out that stars in group two, and even in group three, experienced enrichment linked to massive star cluster evolution.

\section{Early evolution of the Milky Way: clues from galaxy formation models}
\label{sec:model_interpretation}

\subsection{Two regimes of gas accretion onto galaxies and disk formation}
\label{sec:tworegimes}

Although many aspects of galaxy formation are still not fully understood and remain a subject of active research and debate,  studies over the last three decades have helped us to understand the key stages in galaxy evolution. In particular,  two physically distinct regimes of gas accretion onto galaxies were identified: the ``cold flow'' filamentary mode, in which gas accretes along warm filaments that can penetrate to the inner regions of halos and a cooling flow mode wherein gas cools onto galaxy from a hot halo \citep{keres_etal05,keres_etal09,dekel_birnboim06,dekel_etal09}. Physically, the regime of accretion is determined by the ability of a system with a given metallicity and halo mass to maintain a hot halo \citep{birnboim_dekel03}. 

During early stages of evolution, when gas accretion is highly chaotic, progenitors of Milky Way-sized galaxies accrete gas via fast flows along narrow filaments with large angular momentum that is often oriented differently for different flows \citep{stewart_etal11,sales_etal12,danovich_etal15}. This leads to irregular, highly turbulent, extended distribution of star forming gas and stars during early stages of star formation in Milky Way-sized progenitors \citep{rosdahl_blaizot12,stewart_etal13,meng_etal19}.\footnote{ Qualitatively, this picture is a robust prediction of the Cold Dark Matter scenario and is not sensitive to details of galaxy formation physics \citep[e.g.,][]{stewart_etal17}, although whether or not the filaments penetrate all the way to disks may depend on numerical effects and proper treatment of relevant instabilities during supersonic gas flows \citep[e.g.,][]{mandelker_etal19, mandelker_etal20,mandelker_etal21}.}

Stars formed during the early chaotic stages of evolution dominated by cold flow accretion will not be distributed in a thin coherently rotating disk, but in a spheroidal or a very thick disk configuration \citep{obreja_etal13,obreja_etal19,bird_etal13,bird_etal21,yu_etal21}. As shown explicitly by \citet[][]{meng_gnedin21}, for example, 
even if stars are forming in gas with a flattened rotating gas configuration at a given time at high $z$, the disk rapidly changes orientation \citep[see also][]{mccarthy_etal12, tillson_etal15,kretschmer_etal22} or gas distribution changes to a completely non-disk configuration during subsequent epochs.  Likewise, even when stars are forming in a thick perturbed disk at high $z$, the subsequent fluctuations of potential due to mergers and high degree of variability in mass accretion and/or feedback-driven outflows lead to rapid relaxation and spheroidal stellar distribution \citep[][]{kazantzidis_etal08,kazantzidis_etal09,mccarthy_etal12,bird_etal13,dekel_etal20,tamfal_etal21}. Thus, due to the entirety of these factors, during early high-rate accretion stage of evolution most progenitors of the Milky Way mass galaxies cannot form or maintain a thin disk.  

\begin{figure*}
  \centering
   \includegraphics[width=0.9\textwidth]{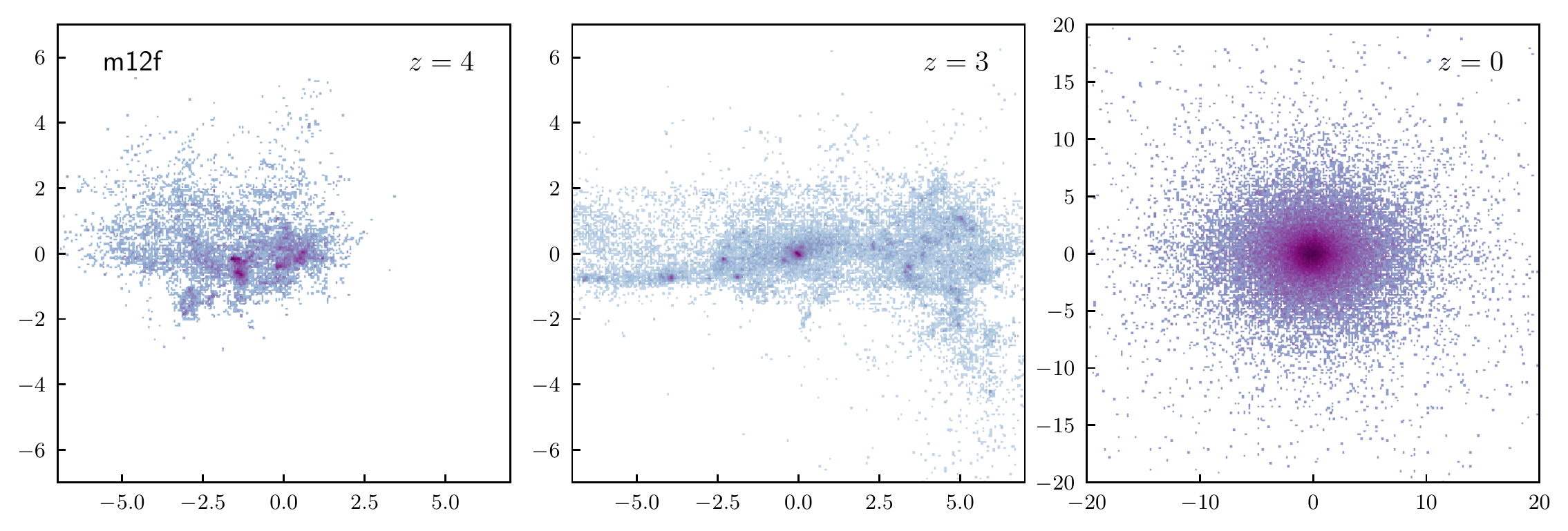}
   \includegraphics[width=0.9\textwidth]{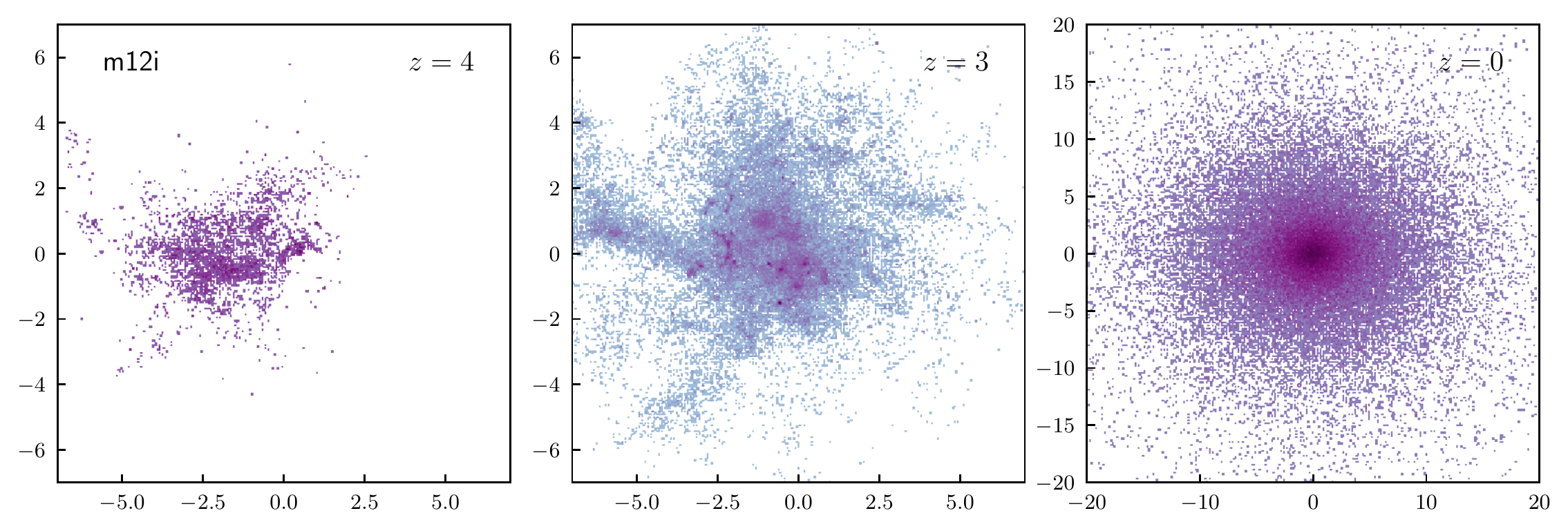}
   \includegraphics[width=0.9\textwidth]{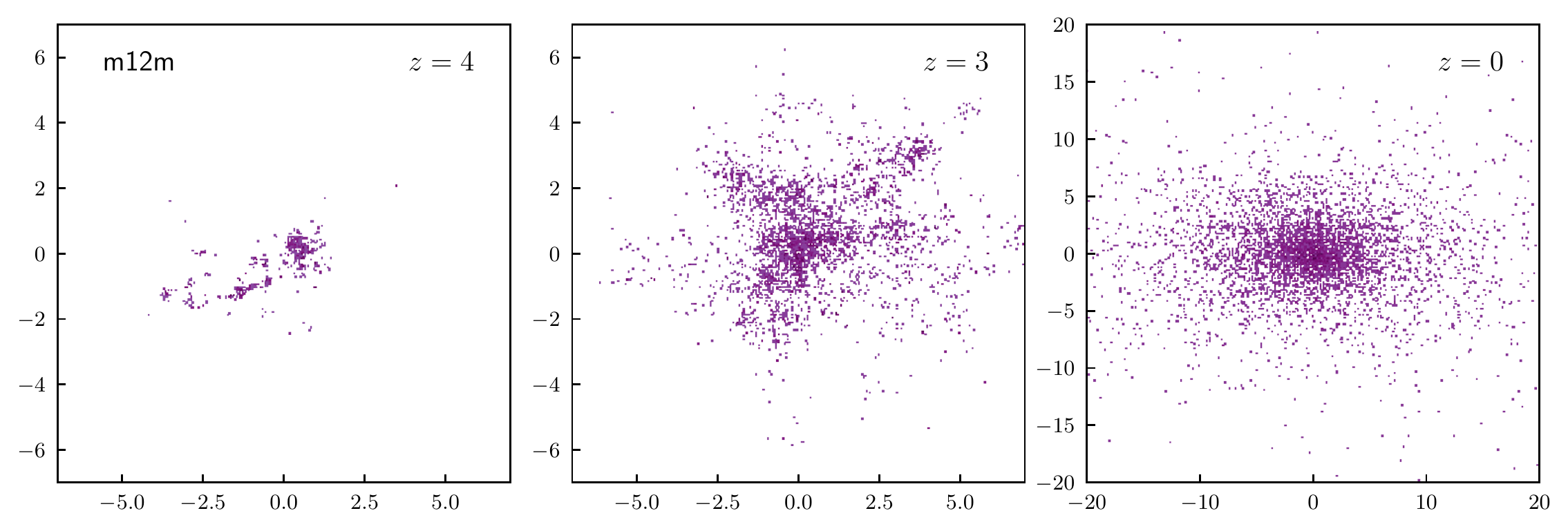}
   \includegraphics[width=0.9\textwidth]{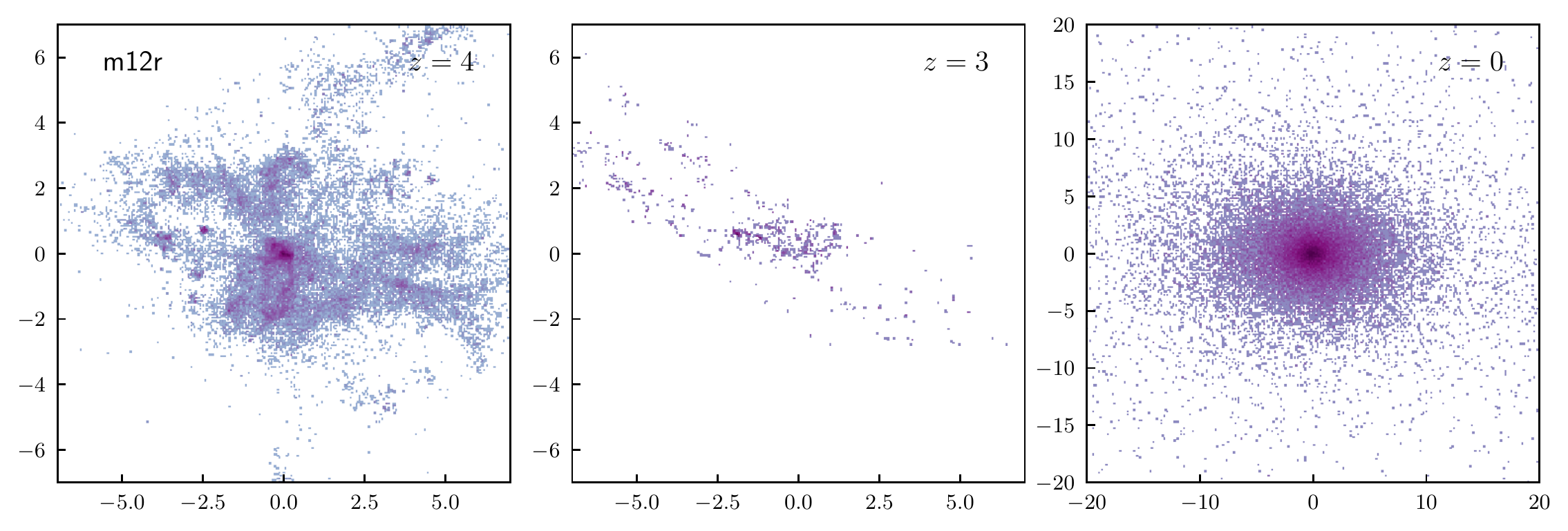}
  \caption[]{Distribution of stellar particles younger than 100 Myr at $z=4$ and $z=3$ ({\it the left two columns in each row}) and distribution of the same particles at $z=0$ ({\it rigtmost panel} in each row) projected in the $X-Z$ plane in four representative MW-sized galaxies from the FIRE-2 Latte simulations suite. The coordinates (indicated on the axes) are in physical kpc; note that the range in the left two columns is the same, but is larger in the rightmost column showing $z=0$ distributions. In each panel the $Z$ axis is aligned with the angular momentum of all stars in the galaxy at the corresponding redshift. The left two columns show a chaotic distribution of young stars forming before the disk formed in each object, while the right column shows that orbital mixing and relaxation result in spheroidal distribution of these stars around galactic center at the present epoch. }
   \label{fig:xz_z}
\end{figure*}

This is illustrated in Figure~\ref{fig:xz_z}, which shows spatial distributions of stellar particles younger than 100 Myr at $z=4$ and $z=3$  and distribution of the same particles at $z=0$ in four representative MW-sized galaxies from the FIRE-2 Latte simulations suite. The figure shows a chaotic distribution of young stars forming at $z=4$ and $z=3$, before the disk appeared in each object, while the right column shows that orbital mixing and relaxation result in spheroidal distribution of these stars around galactic center at the present epoch.

When halo mass and virial temperature become sufficiently large, shock-heated gas does not cool as fast as during early stages and hot gaseous halo can exist \citep{birnboim_dekel03}.  Emergence of such hot halo around the disk results in transition from the early chaotic stage of gas accretion along cold narrow filaments to the slower accretion via cooling of gas from the hot halo or accretion of gas stripped from satellites. This transition is accompanied by ``virialization'' of gas in the inner regions of halo and formation of a coherent long-lived and coherently rotating gaseous disk via a cooling flow \citep{dekel_etal20,stern_etal21}. It is during this stage that the thin stellar disks in spiral galaxies form \citep[][]{hafen_etal22}. The process of transition from the early chaotic distribution of gas to coherently rotating, persistent disk was explicitly studied in the FIRE simulations in a recent study by \citet{gurvich_etal22}. 

The scenario outlined above is consistent with statistical results of \citet{pillepich_etal19}, who used the TNG50 simulation of a $\approx 50$ Mpc box to show that statistically galaxies with disk morphologies in gas and stellar distributions emerge only at redshifts when galaxies have halo masses sufficiently massive to maintain hot halo. For example, their Figures 8 and 9 show that disk morphologies dominate at $z\leq 4$ in galaxies with $M_\star>10^{10}\, M_\odot$, which occupy haloes of mass $M_{\rm h}\gtrsim 5\times 10^{11}\, M_\odot$ \citep[e.g.,][]{behroozi_etal13}. These masses, however, are much higher than the expected mass of MW progenitors at these redshifts. In lower mass haloes, however, distribution of gas and stars is predominantly spheroidal.  

\begin{figure}
  \centering
   \includegraphics[width=0.47\textwidth]{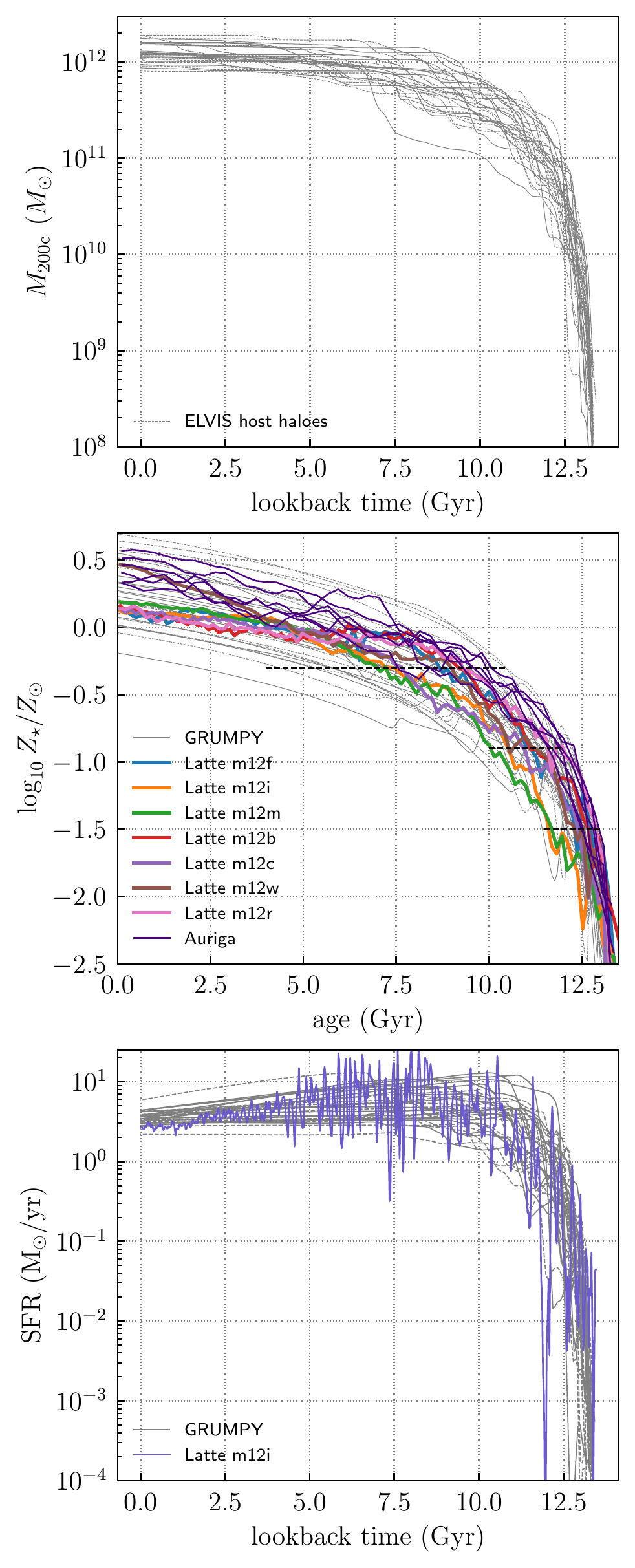}
  \caption[]{Evolution of halo mass defined within the density contrast 200 times the critical density ({\it top panel}), the metallicity--age relation for stars ({\it middle panel}), and evolution of star formation rate ({\it bottom panel}) for a sample of galaxies forming in Milky Way-sized haloes in the GRUMPY regulator model \citep{kravtsov_manwadkar22} ({\it thin gray lines)}, Latte suite of FIRE-2 cosmological zoom-in $N$-body$+$hydrodynamics simulations of three Milky Way-mass galaxies, {\tt m12f}, {\tt m12i}, and {\tt m12m} \citep{wetzel_etal16} ({\it solid indigo lines}). In the middle panel we also show stellar metallicity-age relations in the Auriga simulations of six Milky Way-size haloes \citep{grand_etal18}.}
   \label{fig:host_evo_mod}
\end{figure}
%

\subsection{Evolution of stellar mass, metallicity, and star formation rate in progenitors of MW-sized galaxies}

General implication of galaxy formation models outlined above is that formation of a thin coherently rotating disk in galaxies, such as the Milky Way, is expected to coincide with formation of the inner hot gaseous halo at relatively late stages of evolution $z\lesssim 1-2$. It is thus tempting to identify the transition from the chaotic velocity distribution with $\overline{V}_{\rm tan}\approx 0$ at $\rm [Fe/H]\lesssim -1$ to large $V_{\rm tan}$ at higher metallicities in the observations discussed above with that stage. The metallicity of $\rm [Fe/H]\lesssim -1$, which is lower than the metallicity of the ``Splash'' stars, and so dates from the epochs before the merger with the GS/E galaxy.  
Given that precise stellar ages are not available in observations, we can gauge the typical expected range of ages for stars with $\rm [Fe/H]\lesssim -1$ using theoretical models.

\begin{figure*}
  \centering
   \includegraphics[width=1\textwidth]{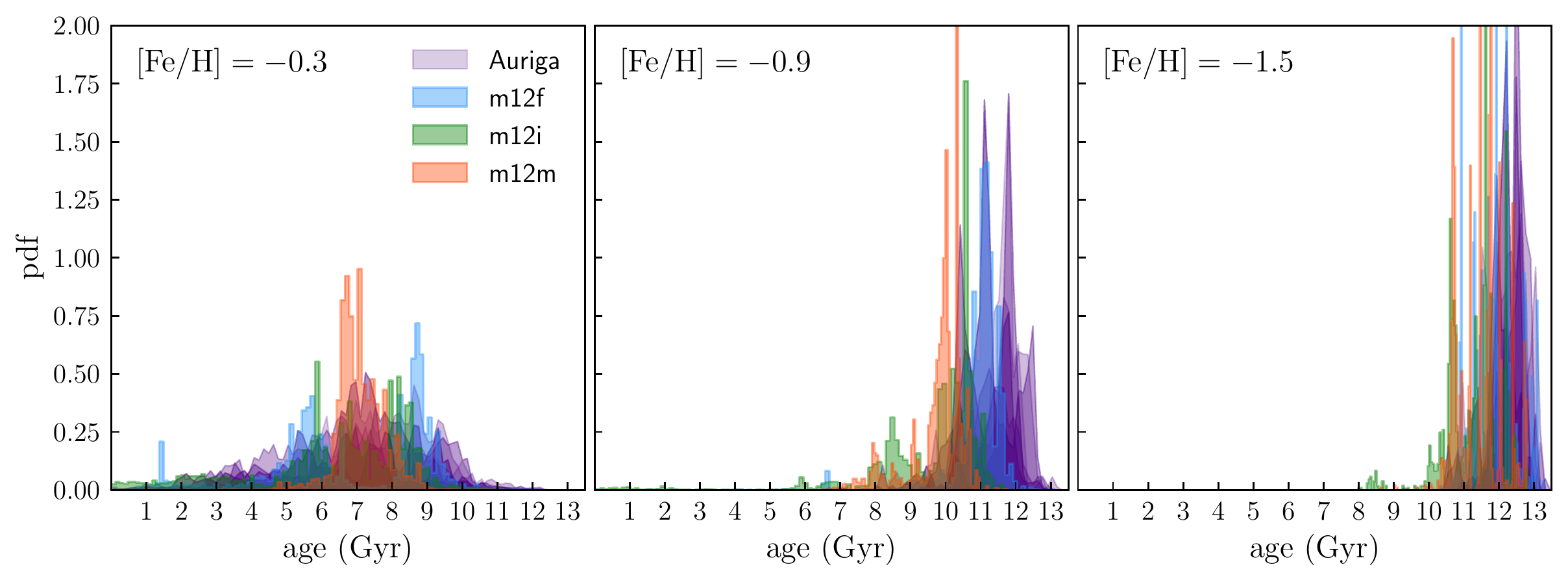}
  \caption[]{Age distributions of stars with metallicities ${\rm [Fe/H]}=-0.3\pm 0.025,\, -0.9\pm 0.025,\, -1.5\pm 0.025$ in the nine simulated MW-mass galaxies from the Latte and Auriga simulations.  In both simulation suites stars are selected within $5<R/{\rm kpc}<11$ and $\vert Z\vert < 3$ kpc.}
   \label{fig:model_age_dist}
\end{figure*}
\begin{figure*}
  \centering
   \includegraphics[width=1\textwidth]{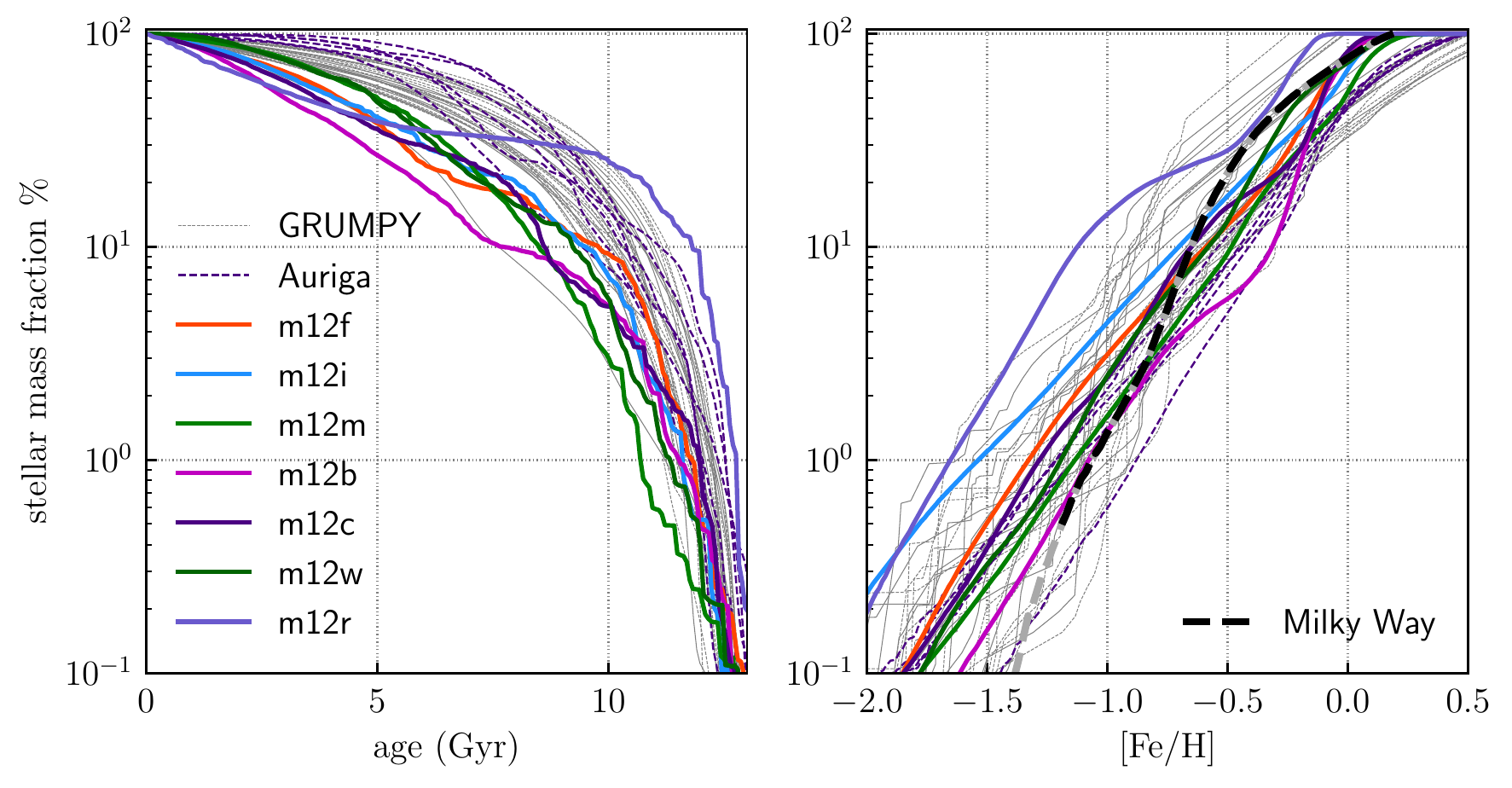}
  \caption[]{Cumulative stellar mass fraction (in per cent) distribution as a function of stellar age ({\it left panel}) and metallicity ({\it right panel}). The {\it thin gray lines} show results of the GRUMPY model for a suite of Milky Way-sized haloes, while the {\it solid colored lines} show results of the Latte suite of the FIRE-2 simulations. The black/gray dashed line shows the estimate of the in-situ stellar mass fraction around the Sun in the Auriga DR17 from the bottom panel of Fig.~\ref{fig:velocity}; the gray portion is at metallicities where a certain confusion of the in-situ and accreted components exists. }
   \label{fig:fstar_cum_model}
\end{figure*}

Figure~\ref{fig:host_evo_mod} shows evolution of halo mass defined within the density contrast 200 times the critical density, as well as the metallicity--age relation for stars and evolution of star formation rate for a sample of galaxies forming in Milky Way-sized haloes in the GRUMPY regulator model \citep{kravtsov_manwadkar22} using mass accretion histories from the ELVIS suite of dark matter simulations \citep{garrison_kimmel_etal14}, and cosmological zoom-in $N$-body$+$hydrodynamics simulations of several Milky Way-mass galaxies from the FIRE-2 simulation public data release\footnote{The Latte simulations were run using the Gizmo gravity plus hydrodynamics code in meshless finite-mass  mode \citep{hopkins15} and the FIRE-2 physics model \citep{hopkins_etal18}.} \citep{wetzel_etal22}.\footnote{\url{http://flathub.flatironinstitute.org/fire}}, as well as the Auriga simulations available in the Aurigaia public release \citep{grand_etal18}.\footnote{\url{https://wwwmpa.mpa-garching.mpg.de/auriga/gaiamock.html}}
Note that we only plot the ``in-situ'' stars in all models: in the GRUMPY model only such stars are modelled for each host, in the FIRE simulations we only use star particles formed at less than 30 comoving kpc from the progenitor centre at each epoch, and in the Auriga simulations we use the in-situ flag provided with the data release to select stellar particles. In both the FIRE and Auriga simulations we select stars with $5<R/{\rm kpc}<11$ and $\vert Z\vert < 3$ kpc. In addition, to calibrate the metallicity distribution to the Milky Way and bring different simulations to a common metallicity scale, we shift stellar metallicities of the selected stars by a constant factor so that their median metallicity equals to the solar value, in agreement with the median metallicity of stars in the Milky Way at these radii \citep[e.g.,][]{katz_etal21}. This shift is small: $\approx 0.05-0.1Z_\odot$ for the Auriga and $\approx 0.01-0.3Z_\odot$ for the FIRE objects (with the most typical shifts of $0.1-0.2$). 

The galaxy evolution models considered here all result in galaxies with realistic properties at $z\approx 0$, but they are very different in how they model galaxy formation. Comparison of their results should therefore give a reasonable estimate of the range of theoretical expectations. 

The two regimes of fast and slow mass accretion separated by the lookback time of $\approx 11$ Gyrs are clearly visible in the evolution of $M_{\rm 200c}$ in the top panel of Figure~\ref{fig:host_evo_mod}. These two regimes are reflected in the two regimes of stellar mass buildup in the lower panel: an extremely fast rise of star formation rate at the lookback times of $\gtrsim 11$ Gyrs and a steady star formation rate at later epochs \citep[on average, cf. also][for analyses of these two regimes in the context of stellar disk formation in the FIRE-2 simulations]{ma_etal17,flores_velazquez_etal21,yu_etal21,gurvich_etal22}. In addition to the overall average behavior of ${\rm SFR}(t)$ in these two regimes, the burstiness of star formation in the FIRE simulations is also very high during early stages of evolution and is much smaller during later stages after virialization of the inner hot halo and formation of the coherent gaseous disk \citep{stern_etal21,yu_etal21,hafen_etal22,gurvich_etal22}, as illustrated by the track of the {\tt m12i} Latte run shown in the lower panel of Figure~\ref{fig:host_evo_mod}. Note that in these simulations burstiness can remain substantial well past the overall transition between the two regimes in the average SFR. For example, in {\tt m12i} this transition occurs at the lookback time of $\approx 10-11$ Gyrs, while star formation remains bursty to $\approx 4$ Gyrs and as a result thin disk in this simulation forms only at the lookback times $\lesssim 3.2$ Gyrs \citep[see Table 1 in][]{yu_etal21}. 

The metallicity of $0.1Z_\odot$, however, is achieved much earlier in the model Milky Way-mass galaxies. Figure~\ref{fig:host_evo_mod} shows that in both the GRUMPY model results and in the Latte and Auriga simulations the stars with $Z\approx 0.1Z_\odot$ typically form at the lookback times $\approx 11\pm 1$ Gyr \citep[see also][]{kruijssen_etal19}, which roughly coincides with the epoch of transition from the fast to slow accretion regime and the corresponding transition from the rapid rise to steady average star formation rate. 

The actual distribution of ages of stars in the Latte and Auriga simulations in the metallicity ranges of ${\rm [Fe/H]}=-0.3\pm 0.025,\, -0.9\pm 0.025,\, -1.5\pm 0.025$ are shown in Figure~\ref{fig:model_age_dist}. The figure shows that the stars of ${\rm [Fe/H]}<-0.9$ are almost all older than 10 Gyr and their distributions have many narrow peaks reflecting strong short-duration bursts of star formation at these early stages of galaxy evolution. The stars of ${\rm [Fe/H]}\approx -0.3$ have a considerably broader age distributions centered on $\approx 6-8$ Gyrs and with much less prominent peaks, which reflects a considerably less bursty star formation at these later epochs.

\begin{figure*}
  \centering
   \includegraphics[width=1\textwidth]{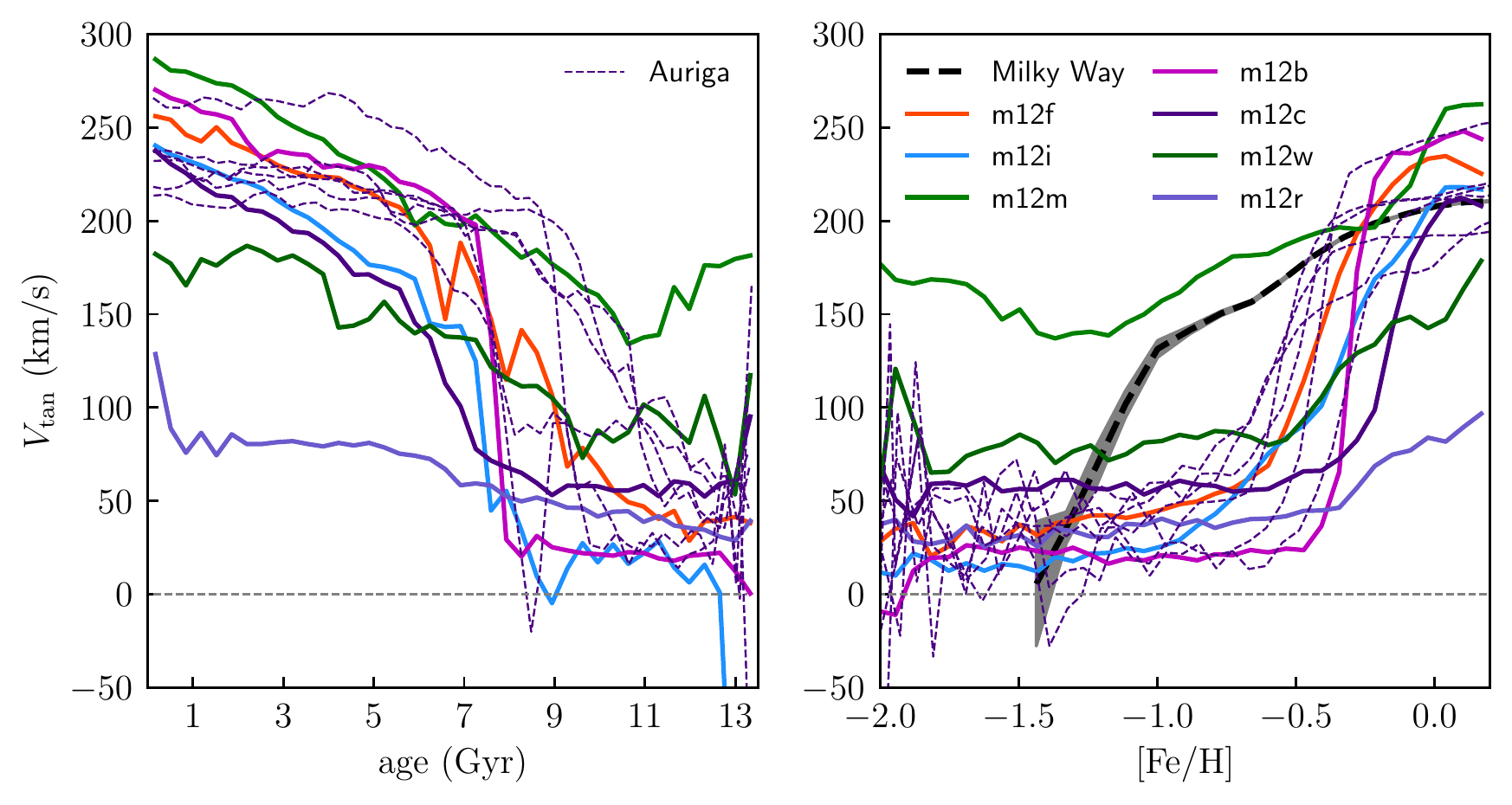}
  \caption[]{Median tangential velocity of stars in the Latte and Auriga simulations as a function of stellar age ({\it left panel}) and metallicity ({\it right panel}). The {\it solid colored lines} show median curves for nine simulated Milky Way-sized galaxies, as indicated in the legend. The {\it black dashed line} shows the median measurement for the Milky Way obtained using the APOGEE DR17 data with shaded gray area representing 68\% uncertainty.}
   \label{fig:vtan_models}
\end{figure*}

The cumulative distributions of stellar mass fraction as a function of age and metallicity are present in Figure~\ref{fig:fstar_cum_model}, which shows that on average $\lesssim 10\%$ of the final mass is formed in model galaxies in the first 3 Gyr of evolution (i.e. $z\gtrsim 2$), which is consistent with observational constraints for the Milky Way-mass galaxies \citep[e.g.,][]{leitner12,van_dokkum_etal13}. The cumulative distribution as a function of metallicity in the right panel of Figure~\ref{fig:fstar_cum_model} is quite similar to the corresponding distribution inferred for the APOGEE data and shown in Figure~\ref{fig:velocity}. In particular,  only $\approx 3\%$ of the final stellar mass in the model galaxies form with metallicities ${\rm [Fe/H]}<-1$, which is close to the value inferred from the APOGEE data.  

\subsection{Transition between star formation regimes and formation of the coherently rotating Milky Way disk}
\label{sec:model_2regimes}

The model predictions shown in Figure~\ref{fig:host_evo_mod} indicate that stars with $\rm [Fe/H]\leq -1$ formed more than 10 Gyrs ago \citep[see also][]{kruijssen_etal19,agertz_etal21}. The observed rapid increase in the median $V_{\rm tan}$ at $\rm [Fe/H]>-1$ thus indicates that a coherent rotating stellar disk formed in the Milky Way progenitor  $\approx 10-11$ Gyr ago (or $z\gtrsim 2.5$). This formation epoch is much earlier than the epochs of disk formation for most of the galaxies in the FIRE-2 suite of simulations \citep{yu_etal21,hafen_etal22,gurvich_etal22}. 

This is also reflected in Figure~\ref{fig:vtan_models}, which shows the median $V_{\rm tan}$ as a function of age and metallicity for stellar particles in the Milky Way--mass galaxies from the Auriga and Latte simulation suites.  Qualitatively, the trends are similar to those of the MW stars shown in Figure~\ref{fig:velocity}: old, low-metallicity stars show little or much smaller mean rotation than younger, higher-metallicity stars. Interestingly, at $\rm [Fe/H]\lesssim -1$ median tangential velocity  in both observations and most of the simulated galaxies is small at $V_{\rm tan}\approx 10-60$ km/s, but it is not zero. Another similarity is that transition from small to large median $V_{\rm tan}$ occurs over a narrow range of metallicities ($\Delta {\rm[Fe/H]}\approx 0.2-0.3$). The trend of $V_{\rm tan}$ as a function of stellar age in the left panel shows that this reflects rapid formation of rotating disks over the time period of $\approx 1-2$ Gyr \citep[see also][]{gurvich_etal22}. 
These qualitative similarities lend credence to the interpretation of the observed trend as signaling formation of the Milky Way disk. At the same time, there are interesting differences.  

\begin{figure*}
  \centering
   \includegraphics[width=0.9\textwidth]{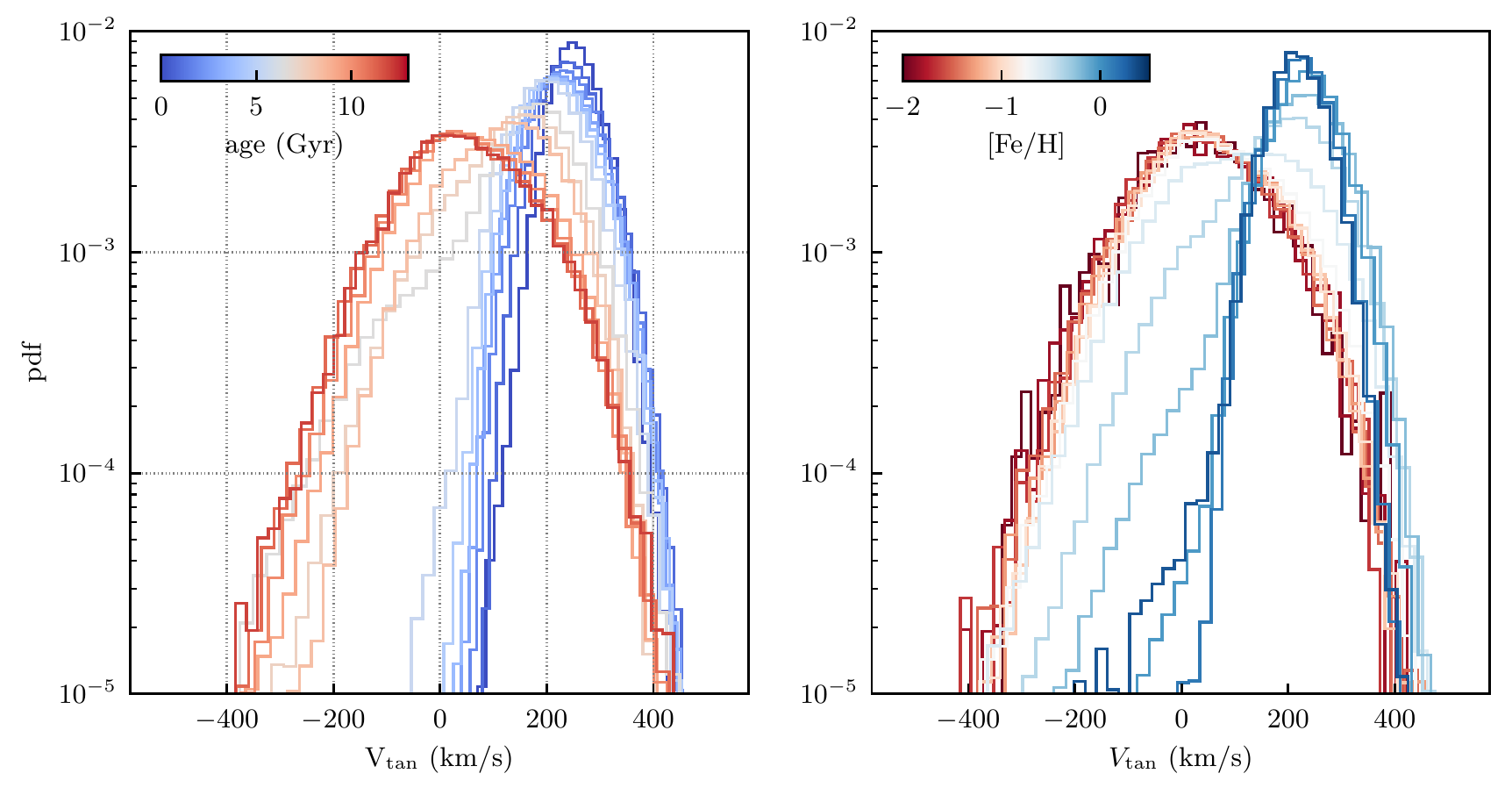}
   \includegraphics[width=0.9\textwidth]{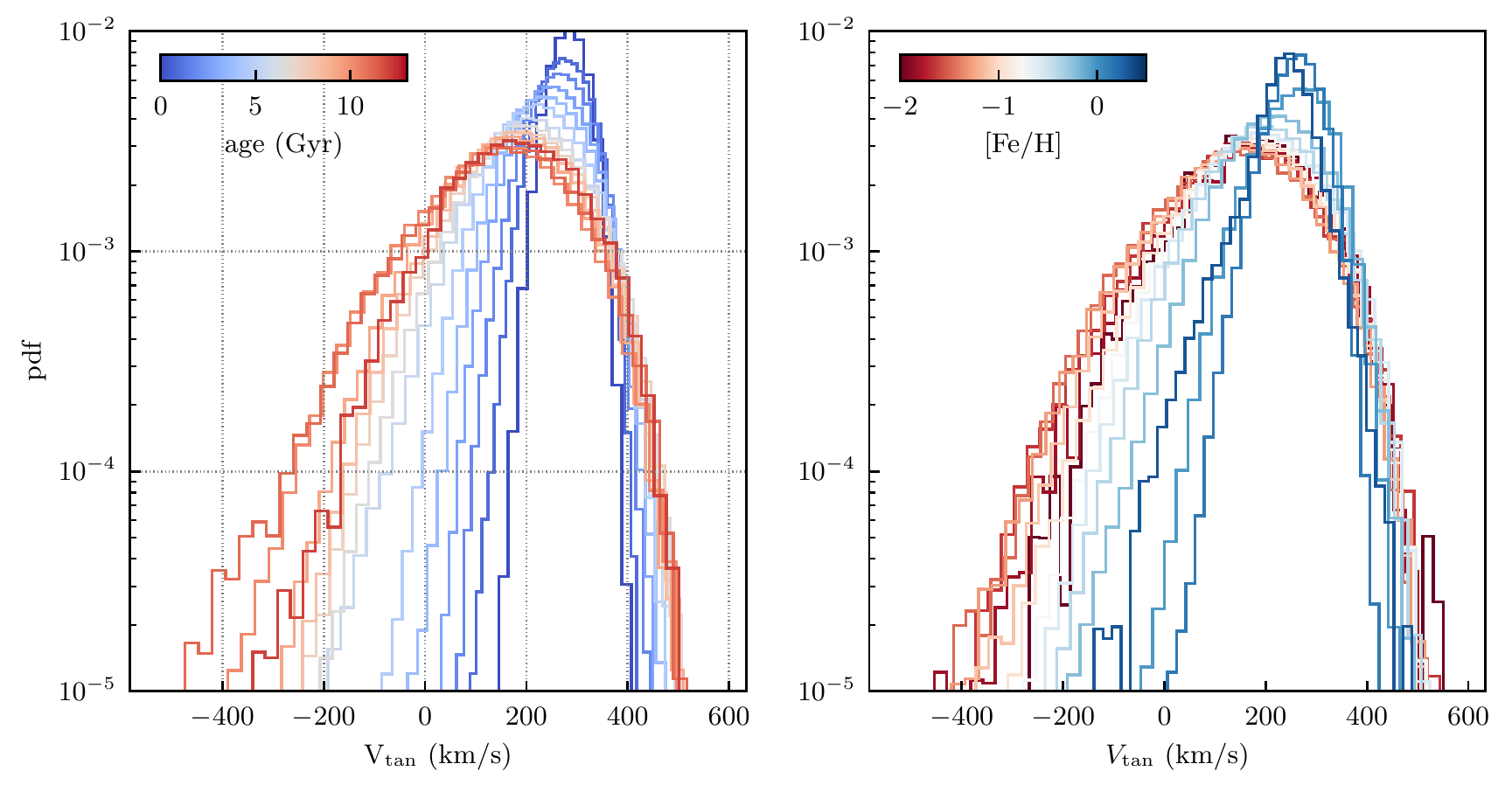}
  \caption[]{Distributions of the tangential velocity, $V_{\rm tan}$, for stars of different ages ({\it left panels}) and metallicities ({\it right panels}) for the MW-mass objects in the {\tt m12f} ({\it top row}) and {\tt m12m} ({\it bottom row}) Latte simulations. The age and metallicity corresponding to each distribution are color-coded as shown in the colorbar in each panel. The distributions are plotted at constant intervals of age and metallicity, so that rapid transition between old, metal-poor distributions in the upper panels to late, metal-rich coherently rotating distributions correspond to rapid transition from the early pre-disk mode of evolution to formation of coherently rotating disk.}
   \label{fig:vtan_dist}
\end{figure*}

The left panel of the Figure shows that transition from small median $V_{\rm tan}$ to value of $\approx 220-250$ km/s occurs in old stars of ages $\approx 7-11$ Gyr. 
The exception is the object {\tt m12m}, in which the mean rotation is present with the median $V_{\rm tan}\approx 100-150$ km/s even in the oldest stars, which indicates that there is a significant variation in the timing of such transitions from object to object \citep[see also][]{yu_etal21}. Indeed, as can be seen in the results of \citet{santistevan_etal20} {\tt m12m} undergoes a significant period of quiescent evolution from $z\approx 5$ to $z\approx 3$, during which its halo mass changed little. Few stars were formed during this period as well. The halo mass then increased tenfold from $z\approx 3$ to $z\approx 2$ and reached the level sufficient to sustain hot halo by the latter epoch leading to formation of a coherently rotating disk where a large fraction of stars formed. 

Figure~\ref{fig:vtan_dist} shows distributions of $V_{\rm tan}$ for stars of different ages and metallicities. In the {\tt m12m} simulation the change of $V_{\rm tan}$ distribution with age is gradual and the form of the $V_{\rm tan}$ distribution of old stars in that simulation is skewed towards positive values, reflecting significant net rotation of old stars in this object. Indeed, even at very low metallicities (${\rm [Fe/H]}$) stars are in a flattened, very thick disk configuration. Evolution of this object is qualitatively consistent with the ``upside-down'' scenario of disk formation \citep{bird_etal13,bird_etal21}, whereby thick disk is forming first while subsequent generations of stars form in colder gaseous disks forming disk of smaller and smaller scale height. Indeed, as shown in Figure~\ref{fig:fire_z_r} in the Appendix, old low-metallicity stars in the {\tt m12m} object are distributed in a very thick, albeit somewhat flattened configuration.  

However, the distribution as a function of age for the MW-mass object in the {\tt m12f} run in the left panel shows two distinct regimes of evolution: the $V_{\rm tan}$ distribution for stars older than $\approx 7$ Gyr is broad and its mode is close to $0\,\rm km\,s^{-1}$, while $V_{\rm tan}$ distribution of younger stars is much narrow and is centered on $\approx 250\,\rm km\,s^{-1}$. The transition between the two forms of distribution is rapid and abrupt for the {\tt m12f} run. The distributions in the two regimes and the rapid transition are also similar in the {\tt m12i}. The ``upside-down'' disk formation in these objects occurs during the last $\approx 6-7$ Gyr, while stars forming at earlier times do not form disk at all, but form a nearly spherical distribution (distribution of low-metallicity, old stars in  the {\tt m12f} object is shown in Figure~\ref{fig:fire_z_r}). 

Formation of the disk in these objects thus does not span their entire evolution history, but is confined to the late stages after the hot inner halo forms. The observations discussed in this study thus probe the transition from the early pre-disk stages evolution of the Milky Way progenitor to formation (``spin-up'') of the Milky Way disk that we inhabit and observe today. Isotropic velocity dispersion measured for low-metallicity observed stars indicate that early stages of MW evolution also formed a spheroidal component, while formation of the disk started about 10-11 Gyrs ago at $z\approx 2-3$. \citet{obreja_etal19} identified a similar spheroidal component in the NIHAO simulations of galaxy formation. They show that this component is kinematically distinct from the bulge, is more spatially extended and is composed almost entirely from the stars formed {\it in-situ}. Moreover, they show that these stars correspond to the earliest episodes of star formation in the Milky Way-sized galaxies. The gas from which these stars form collapses during earliest epochs but loses much of its angular momentum, most likely during collision of gas flowing along different cold streams in the vicinity of the Milky Way progenitor. It therefore forms a spheroidal distribution with little or no net rotation.   

The right panels of Figure~\ref{fig:vtan_dist} show the distribution of $V_{\rm tan}$ for stars of different metallicity and demonstrate that these distribution reveal the two regimes of the star formation similarly to the distributions for age bins, although transition between the two evolution regimes is somewhat smoother due to the scatter between age and metallicity. The overall trend of the distribution shape with stellar metallicity is qualitatively similar to that in the observed stars in the APOGEE survey (see Fig.~\ref{fig:vel_feh}), although the actual shape of the observed low-metallicity stars is not as peaked, while distribution of observed high-metallicity stars is much narrower than in simulations.  

One aspect that is similar in both simulations and observations is that the form of the distributions of $V_{\rm tan}$ in the two regime is quite stable and changes only during narrow range of ages or metallicities in the transition between the two regimes. For young, high-metallicity stars this probably simply reflects the fact that they form in the coherent slowly evolving disk. The similarity of the $V_{\rm tan}$ distribution for old, low-metallicity stars probably reflects the common evolutionary and relaxation processes that these stars have experienced. As discussed in Section~\ref{sec:model_interpretation} above, these processes include star formation in extended irregular gas distribution during cold mode gas accretion and dynamical heating and relaxation due to mergers and overall potential fluctuations due to rapid changes in the mass accretion rate and gas removal in feedback-driven winds.   

Figures~\ref{fig:vtan_dist} and \ref{fig:vtan_models} show that in metallicity transition between the two regimes of evolution is occurring at $\rm [Fe/H]\approx -0.8\div-0.4$ in the simulations, which is considerably larger than in the Milky Way where this transition occurs at $\rm [Fe/H]\approx -1.3\div -1.1$ (see the dashed line). 
Kinematics of the low-metallicity Milky Way stars thus indicates that the virialization of the inner hot halo and subsequent formation of coherently rotating stellar disk occurred quite early in the evolution of our Galaxy and earlier than in most objects in the FIRE-2 and Auriga suites of simulations. Thus, formation of the hot halo and transition from the fast mass accretion regime with highly bursty star formation to slow accretion regime with steady star formation rate  occurred earlier in the Milky Way progenitor. 

Current galaxy formation models differ significantly in the amount of short-term burstiness of star formation rate due to different implementations of star formation and feedback \citep{iyer_etal20} with the FIRE-2 simulations exhibiting the strongest short-term burstiness among the models. The measurements presented in this paper indicate that the epoch when  star formation in the Milky Way progenitor transitioned from the highly bursty to steady regime approximately coincided with the epoch when the {\it average} star formation rate switched from the rapid rise to much slower evolution  $\approx 10-11$ Gyr ago. These measurements can thus potentially serve as a useful constraint on the star formation and feedback physics shaping evolution of the Milky Way mass objects during early stages of their evolution. 

\subsection{Is this scenario consistent with observations of disks in high-$z$ galaxies?}

Observations over the past decade revealed existence of stellar and gaseous disks at $z>3$ \citep[e.g.,][]{stockton_etal08,lelli_etal18,rizzo_etal20,neeleman_etal20,lefevre_etal20,tsukui_iguchi21}. Although interpretation of the Milky Way observations discussed above implies that disk in the Milky Way-mass galaxies forms only at $z\lesssim 2-3$, it does not actually contradict observations of massive high-$z$ disks. This is because these disks are observed in massive, starbursting galaxies with $M_\star\gtrsim 10^{10}\, M_\odot$ and star formation rates $\gtrsim 200-300\, M_\odot\,\rm yr^{-1}$. Such large stellar masses and star formation rates at $z\sim 3-4$ indicate that halos hosting these galaxies have already reached the critical mass threshold required for formation and maintenance of the hot gaseous halo and prevalence of disks in massive high-$z$ galaxies is indeed confirmed by cosmological simulations \citep{feng_etal15,pillepich_etal19}. At the same time, the high star formation rates of these galaxies are inconsistent with the typical rates expected for the Milky Way progenitors \citep[e.g.,][]{behroozi_etal13,behroozi_etal19,moster_etal18}. These massive disks thus probe formation of the progenitors of modern spheroidal galaxies rather than progenitors of the Milky Way, which have not been directly observable at $z\gtrsim 3$ in the deepest observations so far \citep[e.g.,][]{van_dokkum_etal13}. 

Galaxy formation simulations show that flattened, disk-like configurations can arise at high redshifts during evolution of Milky Way progenitors \citep[e.g.,][]{gnedin_kravtsov10,tamfal_etal21}, but such configurations are short-lived at $z>3$, as gas disk orientation generally changes rapidly and overall gas is often not in a disk configuration at all \citep{tillson_etal15,meng_gnedin21,kretschmer_etal22}. 

\begin{figure}
  \centering
   \includegraphics[width=0.48\textwidth]{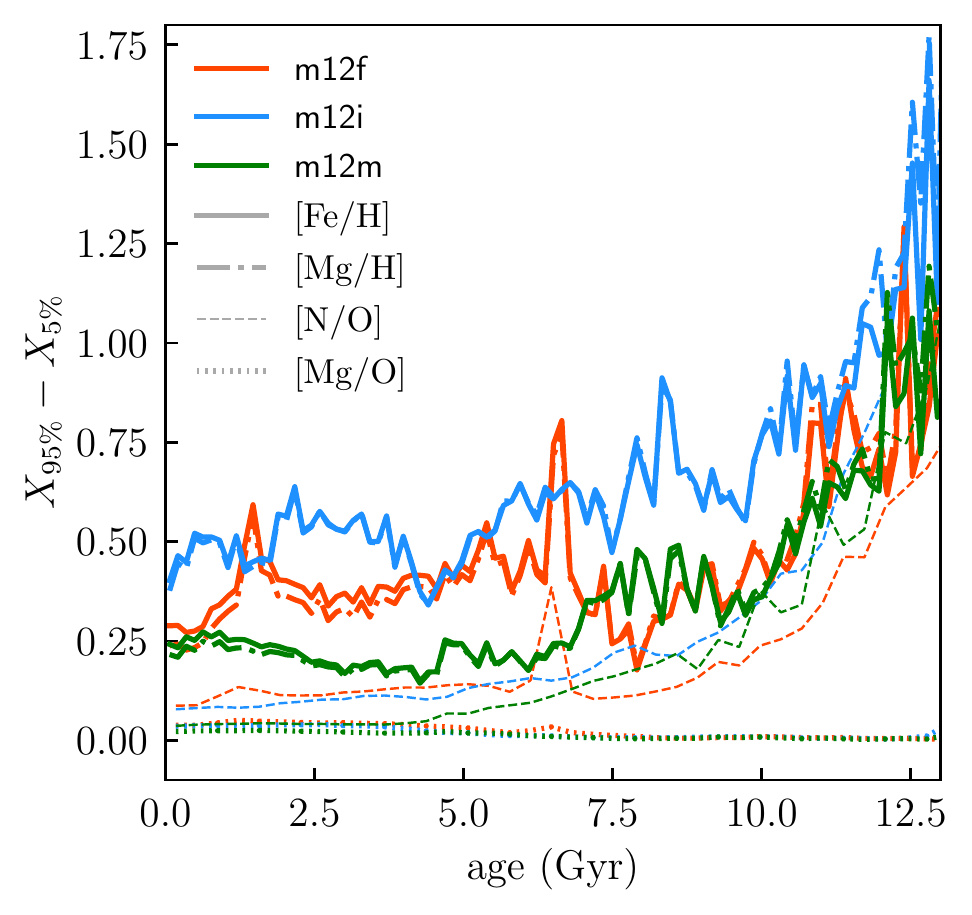}
  \caption[]{$95\%-5\%$ spread in $\rm [Fe/H]$, $\rm [Mg/H]$, $\rm [N/O]$, and $\rm [Mg/O]$ as a function of age of stellar population in the three galaxies from the Latte suite. Rapid variation in star formation, outflow rate, and depletion time during early stages of evoluation (imprinted in the old stars) result in large variations in abundance of all elements (note that the lines for $\rm [Fe/H]$, $\rm [Mg/H]$ are nearly identical). However, the scatter in the element ratios for elements produced by a single source, such as $\rm [Mg/O]$, where both elements are produced exclusively by the core-collapse supernovae, is negligible as both elements vary identically. Scatter in the element ratios, such as $\rm [N/O]$ is, nevertheless, substantial and also sharply increases in old stars, because in this case both SN II and AGB stars contributed significantly to the $N$ abundance, but only SN II contribute to $O$. }
   \label{fig:scatter_age_model}
\end{figure}
%
\subsection{Transition between star formation regimes and evolution of scatter in the element ratios}
\label{sec:chem_scatter_models}

Analyses of the APOGEE DR 17 data presented above showed that the transition from small to large $V_{\rm tan}$ at $\rm [Fe/H]\lesssim -0.9$ is accompanied by a sharp change in scatter in the element abundance ratios $\rm [X/Fe]$ (see Figure~\ref{fig:chem_retro}). Although detailed interpretation of the variation of scatter for different elements is complicated by the uncertainties in theoretical yields \citep[see, e.g.][for a recent review]{matteucci21} and other modelling aspects \citep[see, e.g.,][]{buck_etal21} and is beyond the scope of this study, the increase of scatter at low metallicites is generally consistent with the interpretation that low-metallicity stars have formed during the fast gas accretion regime of the Milky Way formation.

It is well known that abundance of a chemical element depends on the gas accretion rate, gas depletion time, and outflow rates \citep[e.g.,][]{peeples_shankar11}. During the fast mass accretion stage, all of these factors fluctuate strongly both in time and space, which results in the increase of scatter in metallicity. However, if all elements came from the same source, the scatter in metallicity would not result in scatter of the element ratios as all elements would be distributed similarly. To produce the latter scatter, the enrichment history of different elements must be different which, necessitates contribution of multiple sources to their abundance. Moreover, enrichment by the different sources must be different as a function of time and/or metallicity.  In this case fluctuations in star formation rate on the time scales comparable to the enrichment time scales can result in significant scatter in $\rm X/Fe$ \citep[e.g,][]{gilmore_wyse91,johnson_weinberg20}. 

Moreover, fluctuations in the abundance ratios can also arise even when gas mass is constant, but there are strong temporal and/or spatial variations of gas depletion time \citep{weinberg_etal17,johnson_weinberg20}. The density structure of the highly turbulent interstellar medium during early stages of galaxy evolution should be broad and highly variable. The fraction of dense, star forming gas can thus vary rapidly leading to both spatial and temporal variation of local depletion time. Indeed, \citet{meng_etal19} find that local gas depletion times exhibit a wide variation in simulations of high-$z$ progenitors of the MW-sized systems. Large variations in molecular depletion time are also observed in some nearby galaxies with perturbed or unstable gaseous disks and active star formation \citep[e.g.,][]{bolatto_etal11,fisher_etal22}.

Figure~\ref{fig:scatter_age_model} shows the $95\%-5\%$ 
in $\rm [Fe/H]$, $\rm [Mg/H]$, $\rm [N/O]$, and $\rm [Mg/O]$ as a function of age of stellar population in the three galaxies from the Latte suite. Rapid variation in star formation, outflow rate, and depletion time during early stages of evolution (imprinted in the old stars) result in large variations in abundance of all elements (note that the lines for $\rm [Fe/H]$, $\rm [Mg/H]$ are nearly identical), while the scatter in the abundance is approximately constant for stars younger than $10$ Gyr. However, the scatter in the element ratios for elements produced by a single source, such as $\rm [Mg/O]$, where both elements are produced exclusively by the core-collapse supernovae, is negligible as both elements vary identically. Scatter in the element ratios, such as $\rm [N/O]$ is, nevertheless, substantial and also sharply increases in old stars, because in this case both SN II and AGB stars contributed significantly to the $N$ abundance, but only SN II contribute to $O$.

Similarly, results of the NIHAO simulations of Milky Way-sized galaxies presented by \citet{buck_etal21} show that the scatter in element abundance ratios does increase sharply at $\rm [Fe/H]\lesssim -1$ for most enrichment models they considered (see their Figures 9-11). Indicatively, the scatter in this regime is suppressed in the model which assumes that the start of the SN type Ia enrichment is delayed until 158 Myr after the birth of stellar population. This implies that the significant scatter is generated when two sources of chemical elements (core-collapse and type Ia SNe in this case) have different, but temporally close enrichment contributions. 

Likewise, in the Latte simulations we see a sharp increase in the spread of $\rm [N/Fe]$ ratio in stars older than $\approx 10$ Gyr or metallicities $\rm [Fe/H]\lesssim -1$, while in other elements tracked in these simulations the increase is gradual. The distinct feature of nitrogen in these simulations, is that both core-collapse supernovae and AGB stars contribute significantly to its abundance on overlapping time scales, but with rather different time dependence of enrichment by these two sources.  

At metallicities ${\rm [Fe/H]}\lesssim -2$  significant scatter in $\rm X/Fe$ for elements produced by core-collapse supernovae can arise due to strong variations of $\rm X/Fe$ with supernovae progenitor mass \citep{argast_etal00,argast_etal02,scannapieco_etal21}. At such small metallicities enrichment of a local ISM patch can be dominated by a single remnant and remnants of different mass exploding at different times after a star formation burst can result in a wide range of $\rm X/Fe$ when subsequent generation of stars forms from the highly inhomogeneously enriched gas.  

Galaxy formation models indicate that early stages of galaxy formation, where gas-rich, dense and turbulent environments were prevalent and star formation was bursty, are especially conducive  to formation of massive star clusters \citep[e.g.,][]{kravtsov_gnedin05,kruijssen12,kruijssen15,li_etal17,pfeffer_etal18,kim_etal18,ma_etal20}. In particular, the fraction of star formation in bound star clusters was shown to be a function of surface density of star formation both in observations \citep[see, e.g.,][and references therein]{adamo_etal20,adamo_etal20b} and galaxy formation models  \citep{kruijssen12,li_etal17,li_etal22}. Vast majority of massive star clusters that form at high redshifts is expected to dissolve due to high gas mass loss when gas is removed by stellar winds and supernovae \citep[e.g.,][]{li_etal19} and during subsequent evolution due to dynamical heating by strong tides in their natal environment \citep{fall_zhang01,li_gnedin19,li_etal22}. 

These results suggest that a significant fraction of star formation during early stages of the Milky Way formation ($z\gtrsim 2$) occurred in massive star clusters \citep[see, e.g., Fig. 5-8 in][]{pfeffer_etal18}. 
Stars from the massive star clusters formed {\it in-situ} in the Milky Way progenitor and subsequently dissolved could thus contribute substantially to the {\it in-situ} population of disk stars. The self-enrichment during cluster formation, which is manifest in observed star clusters (see Section~\ref{sec:gc_anomalies}), and associated scatter in element abundances (particularly in C, N, O, Al) will thus be manifest in the disk stars originating from disrupted clusters. 

Although the physics of the self-enrichment during cluster formation is not understood \citep[see, e.g., Section 4 of][for a review of models]{bastian_lardo18}, we note that there is now a substantial observational evidence that molecular gas in the natal regions of star clusters is cleared by photoionization and stellar feedback within $\lesssim 3-5$ Myr before the first core-collapse supernovae explode \citep[e.g.,][]{reggiani_etal2011,kos_etal19,kruijssen_etal19b,chevance_etal20,chevance_etal22,kim_etal21}. Such short time frame for star formation in clusters is also strongly indicated by theoretical models \citep[e.g.,][]{li_etal17,li_etal19,semenov_etal21} and implies that models of cluster enrichment that achieve enrichment of the second generation stars with $\approx 3-5$ Myr \citep[e.g.,][see Section 4.6 in \citealt{bastian_lardo18} for a review]{prantzos_charbonnel06} are preferred. 

Detailed modelling of the element abundance ratios and their scatter may provide a unique window into the contribution of different sources to the abundances of a given element and associated enrichment timescales.

\section{Comparison with previous works}

\name, the new in-situ spheroidal component revealed in the APOGEE DR17 and Gaia EDR3 data in this work has previously eluded identification, primarily due to the lack of detailed elemental abundances that would allow to differentiate the in-situ and the accreted halo stars at low [Fe/H]. Initially, pre-{\it Gaia} analyses pointed out the existence of a broad-brush dichotomy in the stellar halo's chemo-kinematic properties \citep[see][]{Chiba2000,Chiba2001,Carollo2007,Carollo2010}. According to these authors, the more metal-rich (with typical metallicities around $-1.5<$[Fe/H]$<-1$) and slightly prograde inner stellar halo was distinct from the more metal-poor (with [Fe/H]$<-2$) and possibly retrograde outer stellar halo \citep[see also][]{Carollo2012,Beers2012, An2013,Santucci2015}. 

{\it Gaia} astrometry revealed that the strong dependence of the stellar halo kinematics on metallicity was due to the superposition of two accreted components, i.e. the tidal debris from a single massive ancient merger at [Fe/H]$\approx-1.3$, i.e. the GS/E material, and contributions from multiple lower-mass accretion events at lower [Fe/H] \citep[see e.g.][]{Belokurov2018, Deason2018, Helmi2018, Mackereth2019}. Note that the first glimpses of the GS/E tidal debris cloud had been seen earlier in \citet{Brook2003}, while the dominance of the single massive ancient merger across the inner halo was first established in \citet{Deason2013}.

In the last decade, an additional stellar halo component containing stars formed in-situ in the Milky Way progenitor was identified using precise elemental abundances \citep{Nissen2010,Hawkins2015}. However, due to relatively small sample sizes used in these studies they probed only metallicities of [Fe/H]$\gtrsim -1$. As discussed in Section~\ref{sec:ymw}, most stars in this metallicity range are in a coherently rotating disk and thus have high tangential velocities, but there is a noticeable tail reaching small and retrograde $V_{\rm tan}$. The latter stars have spheroidal distribution, but were formed in the disk and subsequently dynamically heated during the GS/E merger.

The {\it Gaia} data was instrumental in confirming the genesis of this in-situ halo component, also called the Splash \citep[see e.g.][]{Bonaca2017, Gallart2019, Dimatteo2019, Belokurov2020}. While in the Solar neighborhood the Splash can contribute as much as a half of all stars on halo-like orbits, it is significantly less extended than the accreted halo \citep[see e.g.][]{Belokurov2020,Iorio2021}. The chemical properties of the Splash stars are similar (or often indistinguishable) to those of the thick disk \citep[][]{Nissen2010,Fernandez2018,Fernandez2019,Belokurov2020,Kordopatis2020}, in agreement with our analysis. Similar to numerical merger simulations, the Milky Way's Splash contains a stellar population on average younger compared to the GS/E debris with a tail of ancient stars \citep[][]{Belokurov2020,Bonaca2020} however some disagreement still exists as to the exact star-formation histories of the Splash and the GS/E \citep[c.f.][]{Gallart2019,Montablan2021}. 

In this study, where the sample size allows us to explore in-situ stars with metallicities of [Fe/H]$< -1$, we showed that these low-metallicity stars constitute a new component, which we called the {\name}. These stars have nearly isotropic velocity ellipsoid, little net rotation, and wide distribution of tangential velocities. They also exhibit element abundance ratios distinct from both the accreted halo and the heated disk component. First, the in-situ and the accreted stars follow different tracks in [X/Fe] vs [Fe/H] space. Second, neither GS/E nor Splash show an excess of the abundance dispersion in Al, Si, N and O. 

Note that \namesp stars do not contribute significantly to the Splash population. This is because the heating action of the merging massive satellite is reduced if the velocity dispersion of the host stars is high to begin with and cannot be easily increased much further \citep[see, e.g.,][]{Grand2020}.

We believe that some of the stars with the largest prograde $V_{\rm tan}$ were previously hypothesised to belong to an ancient metal-weak disk  \citep[e.g.,][]{Chiba2000,Beers2002,Carollo2010,Kordopatis2013,Li2017,Sestito2019,Dimatteo2020,Fernandez2021}.  
Here, instead of splitting stars into halo and disk, we show that the low-metallicity in-situ population has a broad, assymetric tangential velocity distribution with the mode that is positive but small ($V_{\rm tan}\approx50$ km s$^{-1}$). The tail of the prograde velocities extends to $V_{\rm tan}\approx 300\,\rm km\,s^{-1}$, while the tail of retrograde velocities extends to $-200\,\rm km\,s^{-1}$. The positive mode and the asymmetry create an overabundance of stars with large prograde velocities, but the distribution is continuous and does not show a clear evidence for a distinct disk component. Instead, we argue that the in-situ spheroidal component with [Fe/H]$<-0.9$ is the phase-mixed remnant population of the early turbulent stages of the Milky Way evolution. Note, however, that orbits of some stars that are born in a spheroidal pre-disk component will evolve as the disk grows and as a result the properties of those stars orbiting close to the Galactic plane will appear more disky at the present day \citep[see e.g.][]{Binney1986}. 

In parallel to the observational studies of the accreted tidal debris, components of stellar halo and their origin were studied using numerical simulations. Nevertheless, to the best of our knowledge the \namesp component revealed here has not been previously identified in simulations. A number of studies explored in-situ halo \citep[][]{Font2006,Zolotov2009,Zolotov2010,purcell_etal10,Font2011,mccarthy_etal12,Tissera2012,Tissera2013,Tissera2014,Cooper2015,Pillepich2015}, but its main formation mechanism was argued to be dynamical heating and dispersing of the stars born in the inner Galaxy. Often these stars originate i) inside a (proto)disk and ii) slightly later compared to the stars in the small-mass dwarfs \citep[e.g.][]{Font2011,mccarthy_etal12,purcell_etal10}. Additionally, in many cases, stars formed from the gas accreted from the merging dwarf satellites are attributed to the in-situ population \citep[see][]{Tissera2012, Cooper2015, Pillepich2015}. Sometimes, stars can form directly out of the hot gaseous halo around the host galaxy, however this mode of in-situ star-formation is suspected to be a numerical artefact \citep[see][]{Cooper2015}. 

In contrast, as was discussed in detail in the previous section,  \namesp forms sufficiently early, before any coherent disk is established. Additionally, while the high-redshift prolific accretion activity probably helped to perturb and displace the low-metallicity stars, we stress that the \namesp stars were {\it born} in a chaotic state, inherited from the highly turbulent and spatially irregular Galactic ISM at the time.  

\section{Summary and Conclusions}
\label{sec:conclusions}

Taking advantage of the exquisite measurements of the element abundances in the APOGEE DR 17 and the {\it Gaia} EDR3 astrometry, we separate low metallicity stars with [Fe/H]\,$\lesssim-0.9$ into the stars born in the Milky Way proper ({\it in-situ}) and the stars created in low-mass dwarf galaxies and later {\it accreted} onto the Milky Way using the [Al/Fe] ratio, as first pointed out by \citet{Hawkins2015}. The method exploits the fact that production of aluminium in dwarf galaxies is suppressed and delayed compared to the rapid and prolific enrichment taking place in the early Milky Way. As a result, our low-metallicity in-situ sample exhibits negligible contamination by accreted stars and provides the first clear view of the state of our Galaxy's progenitor at high redshifts before and during formation of the disk. Our main results and conclusions are as follows. 

\begin{itemize}
    
\item[(i)] We show that the low-metallicity ([Fe/H]$\lesssim -1.3$) in-situ component is kinematically hot with an approximately isotropic velocity ellipsoid and a modest net spin (see Figures~\ref{fig:velocity}, \ref{fig:vel_feh} and Section~\ref{sec:kinematics}). We use galaxy formation models to establish that stars in this component, which we dub \name, reflect the chaotic early pre-disk stages of Milky Way's evolution when the first few per cent of its stars form in highly irregular spatial and velocity configurations and which later phase-mix into a spheroidal distribution (Section~\ref{sec:tworegimes} and Fig. \ref{fig:xz_z}).   

\item[(ii)] We demonstrate that the low-metallicity stars of the \namesp component exhibit a large scatter in element abundances (Section~\ref{sec:chemistry} and e.g. Figures~\ref{fig:chem_stat}, \ref{fig:xmg_stats}). By comparing tracks of individual elements referenced to either Fe or Mg we show that in the early Milky Way, across a variety of chemical elements, the observed scatter of the element ratios is likely caused by the increased stochasticity in metallicity at early times driven by strong variations in gas accretion and gas outflow rates and associated burstiness of star formation (Section~\ref{sec:chem_scatter_models}). 

\item[(iii)] In addition, Al, Si, N and O ratios to either Fe or Mg display an even larger scatter at low metallicities. Comparisons with the anomalous chemical patterns observed in the Milky Way globular clusters indicates that nucleo-synthetic channels acting inside massive stellar agglomerations may have contributed significantly to the overall enrichment of the early Milky Way (Section~\ref{sec:gc_anomalies}, Figure~\ref{fig:al_si_n}).

\item[(iv)] The median tangential velocity of the in-situ stars increases sharply with increasing metallicity between [Fe/H]$=-1.3$ and $-0.9$ (Section \ref{sec:kinematics} and Fig.~\ref{fig:velocity}). The observed and theoretically expected age-metallicity correlations imply that this increase reflects a rapid formation of the Milky Way disk over $\approx 1-2$ Gyrs, during which it {\it spins up} and settles into a coherently rotating, thick disk with a median azimuthal velocity of $V_{\rm tan}\approx 150$ km s$^{-1}$ at ${\rm [Fe/H]}\approx -0.9$. 

\item[(v)] Observations and theoretical models indicate that violent head-on collision with the GS/E dwarf galaxy, which dramatically transformed some of the disk stellar orbits, as manifested in the Splash stellar component, has occurred {\it after} formation of the thick disk in the MW's progenitor.

\end{itemize}

In galaxy formation models gas distribution in the central regions of young Milky Way analogs is highly irregular, turbulent and rapidly evolving due to a combined effects of supersonic cold filamentary flows and feedback-driven winds. Stars born in the densest regions of this gas inherit its irregular distribution and stellar component of most progenitors of the MW-sized galaxies thus does not develop a persistent rotating disk during these epochs. Recent analyses of galaxy formation simulations showed that formation of the disk starts when galaxy is able to build up and sustain hot gaseous halo, which changes the mode of gas accretion and allows development of coherently rotating disk. The thick disk likely forms during this transition, while the subsequent evolution when gas is accretted via a steady cooling flow leads to a stable disk orientation and build-up of the thin disk. 

Most importantly, we stress that the absolute majority, i.e. 12 out of 13 of the numerical simulations studied here exhibit qualitatively similar chemo-kinematic behaviour observed in the Milky Way. Specifically, at the metallicities corresponding to the lookback times between 8 and 12 Gyrs, most model Milky Way analogues have low-amplitude net spin but no coherent disk. The {\it spin-up} transition uncovered in this work is a ubiquitous feature in these models with the young Milky Ways in both Auriga and FIRE simulations undergoing a fast transition from a messy slowly-rotating \name-like stellar component to a fast rotating disk. 

Interestingly, in most of these models the {\it spin-up} transition happens at significantly larger metallicities than observed in the Milky Way. If we assume that the spin-up observed in the Milky Way is typical, there may be several possible reasons for this difference. First, note that the low-metallicity in-situ stars probing this transition constitute only a few per cent of the final  stellar mass (see Figs.~\ref{fig:velocity} and \ref{fig:fstar_cum_model}). To reproduce the observed spin-up metallicity simulations must correctly model both the early star formation history and chemical enrichment of this component (including removal of heavy elements by winds) and thermodynamical processes related to the formation of hot gaseous halo and related dynamical processes leading to disk formation. Given these considerations, discrepancy with current generation of simulations is not surprising. At the same time, going forward the observed spin-up will provide a sensitive probe of the physics of these processes and their treatment in galaxy formation simulations.


We use galaxy formation simulations and results of chemical enrichment models to show that the elevated scatter in stellar metallicity and abundance ratios is driven by strong bursts of star formation during early pre-disk stages of evolution (see Section~\ref{sec:chem_scatter_models}). The bursts are often separated by gaps in star formation activity. The corresponding gaseous outflows and ongoing fresh gas accretion result in large scatter in element abundances. In particular, we show that low-metallicity stars born in the FIRE-2 suite of simulations clearly display rapid increase of metallicity dispersion with decreasing metallicity for stars born during the pre-disk phase. 

Additionally, we use simulations to explicitly show that scatter in chemical abundance ratios can be pushed up when multiple stellar sources contribute to abundance of a given chemical element on different time-scales, while scatter in the abundance ratios are negligible if the two elements in the ratio are produced by the same single source. While multiple sources are thought to contribute to the abundance of most chemical species, details of these processes are uncertain and their modelling in the current simulations of galaxy formation and chemical enrichment is not yet adequate for a direct detailed comparisons with observations trends. 

Results of this study indicate that the low metallicity tail of Milky Way's stellar population provides a unique window into the tumultuous state of the Galaxy in its pre-disk and disk formation stages. This may be our only view of the detailed evolution of Milky Way-sized galaxies as their expected stellar masses and star formation rates are too low to be directly observable at $z\gtrsim 2$. Future spectroscopic surveys such as DESI, WEAVE, 4MOST and SDSS V will significantly increase the size of the low-metallicity sample and this should provide multiple exciting avenues for exploration. 

\section*{Acknowledgments}

The authors are grateful to Keith Hawkins, Kathryn Johnston, Thomas Masseron, Julianne Dalcanton, David Hogg, Adrian Price-Whelan, Zephyr Penoyre, David Aguado, Daniel Angl\'es-Alc\'azar, Drummond Fielding, Wyn Evans, Andreea Font and the members of the Cambridge Streams and UChicago Friday Owls Clubs for many enlightening conversations that helped to improve the quality of this work. We are also grateful to Claude-Andr\'e Faucher-Giguere and Jonathan Stern for comments and sharing their paper (Gurvich et al.) ahead of publication. AK would also like to thank Institute of Astronomy at Cambridge University and its Sackler visitor program for the warm hospitality during his 2018-2019 visit. His work on this project was supported by the National Science Foundation grants AST-1714658 and AST-1911111 and NASA ATP grant 80NSSC20K0512.

In this study we have used  the Ananke database of the Latte simulations (\url{http://fire.northwestern.edu/ananke/}), FIRE-2 simulation public data (\url{http://flathub.flatironinstitute.org/fire}), and Auriga (\url{http://wwwmpa.mpa-garching.mpg.de/auriga/}) simulation teams for making their simulation results publicly available. We use simulations from the FIRE-2 public data release \citep{wetzel_etal22}. The FIRE-2 cosmological zoom-in simulations of galaxy formation are part of the Feedback In Realistic Environments (FIRE) project, generated using the Gizmo code \citep{hopkins15} and the FIRE-2 physics model \citep{hopkins_etal18}. 

This research made use of data from the European Space Agency mission Gaia
(\url{http://www.cosmos.esa.int/gaia}), processed by the Gaia Data
Processing and Analysis Consortium (DPAC,
\url{http://www.cosmos.esa.int/web/gaia/dpac/consortium}). Funding for the
DPAC has been provided by national institutions, in particular the
institutions participating in the Gaia Multilateral Agreement. This
paper made used of the Whole Sky Database (wsdb) created by Sergey
Koposov and maintained at the Institute of Astronomy, Cambridge with
financial support from the Science \& Technology Facilities Council
(STFC) and the European Research Council (ERC). 

\bibliography{references}

\appendix
\section{Distribution of stars in the FIRE simulations}
\label{sec:fire_r_z}

\begin{figure*}
  \centering
   {\includegraphics[width=0.49\textwidth]{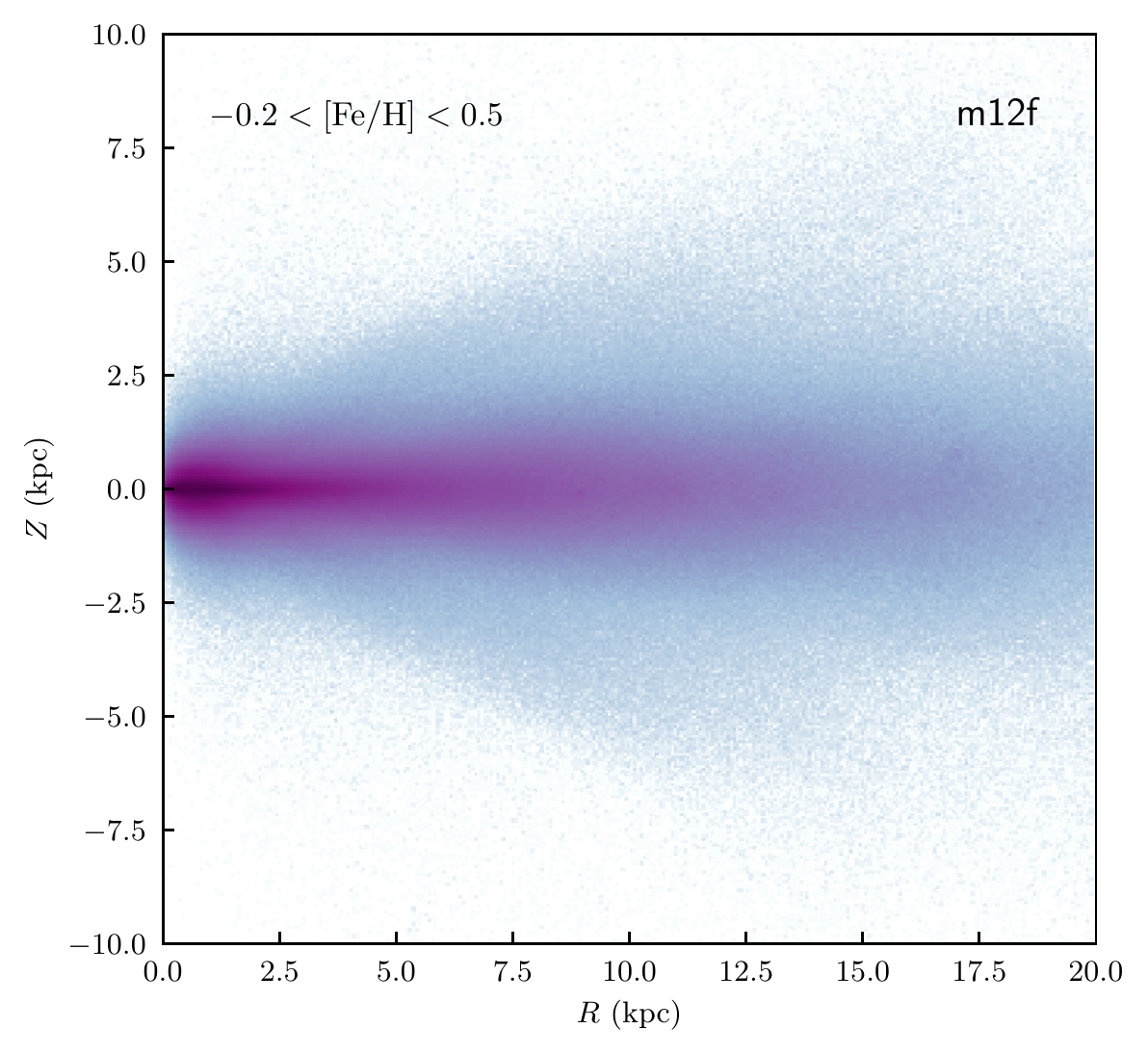}
   \includegraphics[width=0.49\textwidth]{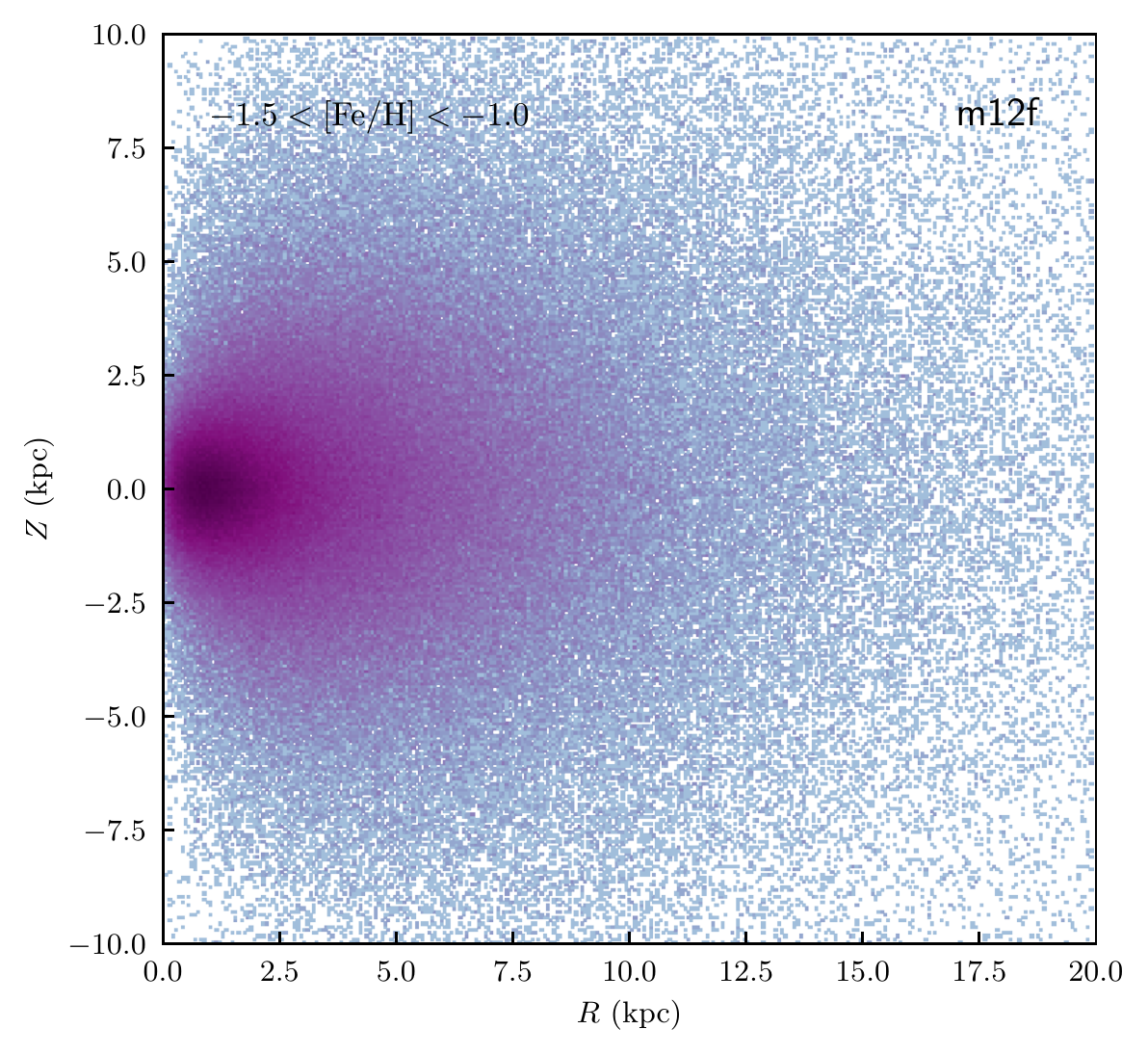}}
   {\includegraphics[width=0.49\textwidth]{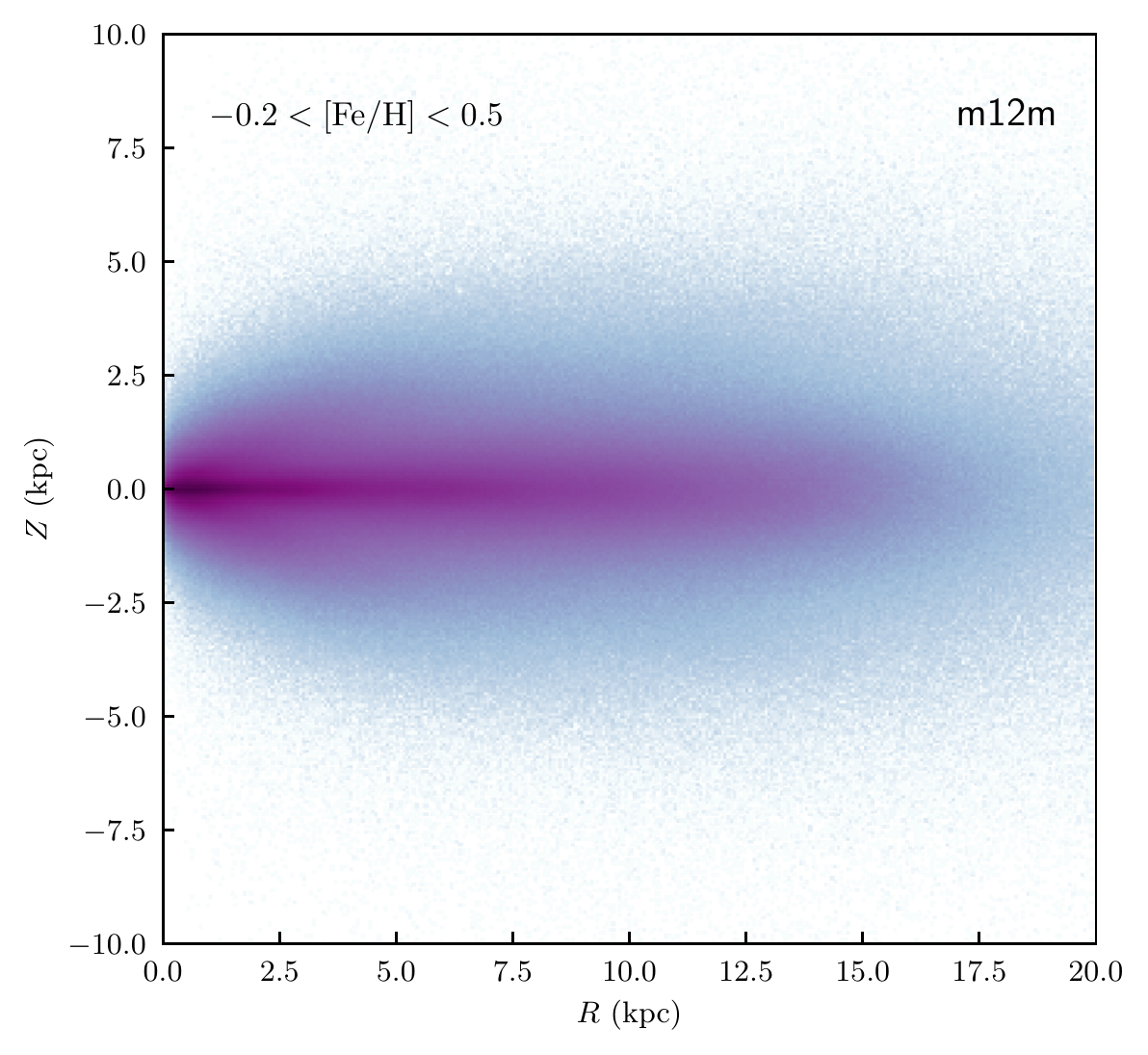}
   \includegraphics[width=0.49\textwidth]{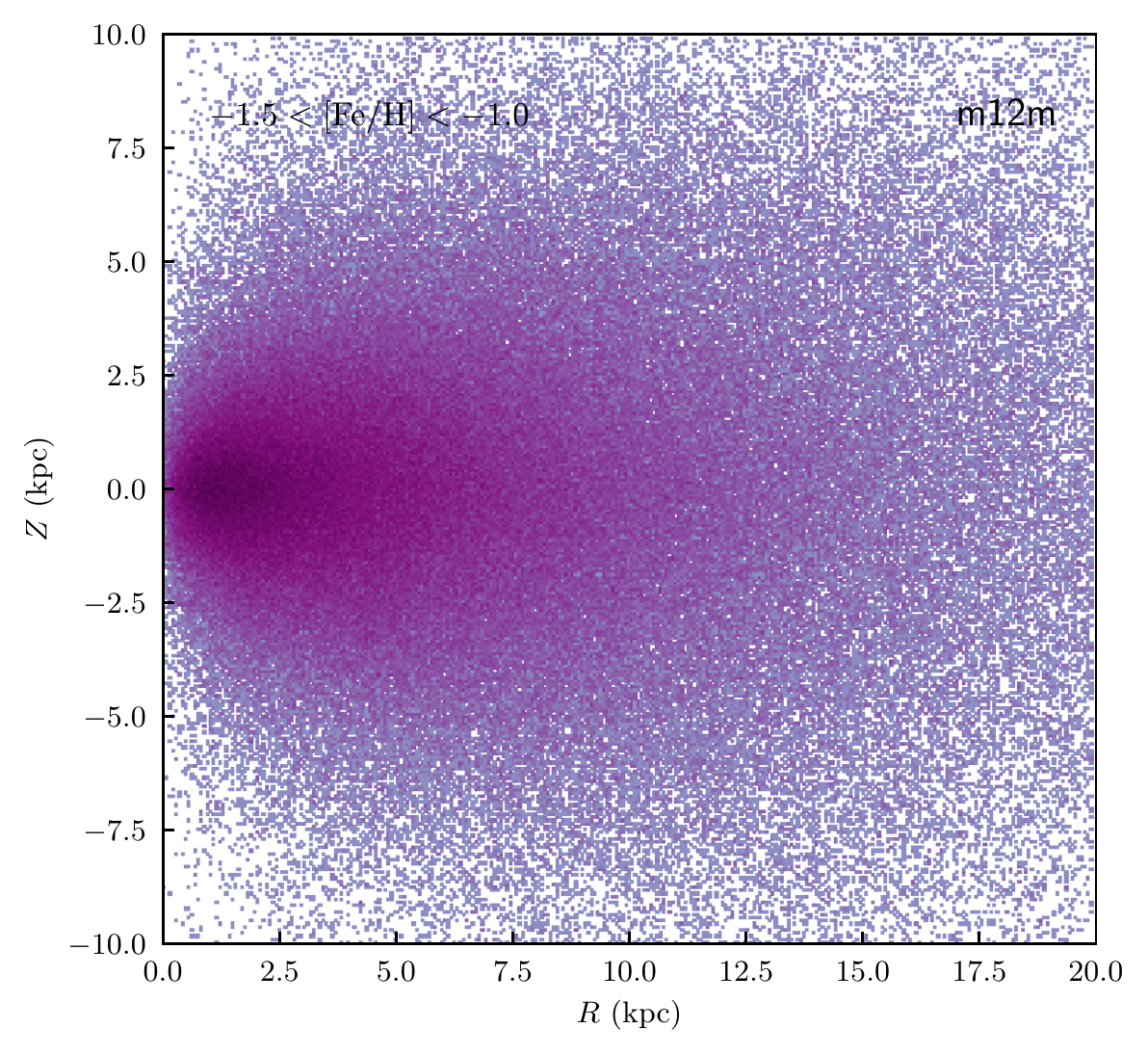}}
  \caption[]{Distribution of the in-situ stellar particles at $z=0$ of two MW-sized galaxies from the FIRE suite in the $Z-R$ plane, where $Z$ is the distance from the plane of the disk defined as the plane perpendicular to the angular momentum of all stars and $R$ is the galactocentric distance. The {\it top row} shows  {\tt m12f} object  and {\it bottom row} shows {\tt m12m}; the left panels show stars with $-0.2<{\rm [Fe/H]}<0.5$ that form in the last several Gyrs and are distributed in a disky configuration, while the right panels show metal poor stars with $-1.5<{\rm [Fe/H]}<-1$ that form during earlier chaotic epochs which have a spheroidal distribution at the shown $z=0$ epoch.} 
   \label{fig:fire_z_r}
\end{figure*}

Figure~\ref{fig:fire_z_r} shows distribution of the in-situ stellar particles at $z=0$ of two MW-sized galaxies from the FIRE suite ({\tt m12f} and {\tt m12m}) in the $Z-R$ plane, where $Z$ is the distance from the plane of the disk defined as the plane perpendicular to the angular momentum of all stars and $R$ is the galactocentric distance. The figure shows stars with $-0.2<{\rm [Fe/H]}<0.5$ (left panels) that form in the last several Gyrs and are distributed in a disky configuration and metal poor stars with $-1.5<{\rm [Fe/H]}<-1$ (right panels) that form during earlier chaotic epochs. The figure shows that low-metallicity stars have a spheroidal distribution at the shown $z=0$ epoch, even in {\tt m12m}, in which stars forming at $z>2$ exhibited coherent rotation (although they did not form in a well-defined disk, see Fig.~\ref{fig:xz_z}).

\section{Asymmetry of the chemical abundance distribution at [Fe/H]$<-0.9$}
\label{sec:asymmetry}

\begin{figure*}
  \centering
   \includegraphics[width=0.99\textwidth]{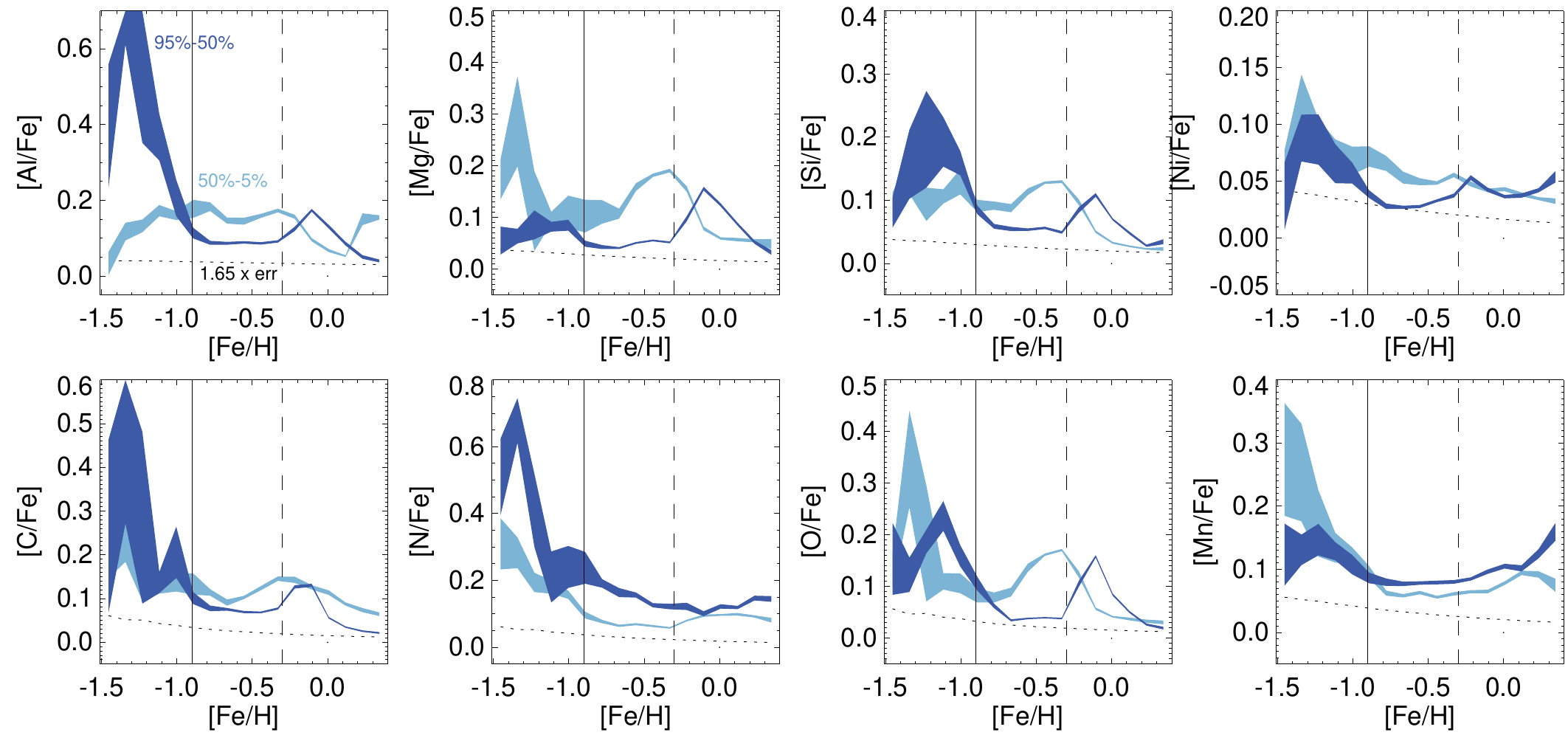}
  \caption[]{Energy vs $L_{\rm z}$.}
   \label{fig:chem_stat_pm}
\end{figure*}

Figure~\ref{fig:chem_stat_pm} gives $+1.65\sigma_{\rm[X/Fe]}$ ($-1.65\sigma_{\rm[X/Fe]}$) measured as the difference between the 95th and 5th (50th and 5th) percentiles of [X/Fe] distribution in dark (light) blue. These can be compared to $2\times$ the uncertainty of the abundance ratio measurement as a function of [Fe/H] shown as a dotted line in each panel. Interestingly, for several elements there exist ranges of [Fe/H] where the measured one-sided spread is as narrow as the measurement uncertainty. This confirms that while astonishingly low, the ADR17 abundance uncertainties nonetheless give a good representation of the quality of the measurement. At [Fe/H]$>-0.9$ the elements where a clear bifurcation at fixed metallicity is present, the $+1.65\sigma$ and $-1.65\sigma$ curve exhibit distinct behaviour as the prominence of the upper and lower branches changes with [Fe/H]. For [Fe/H]$<-0.3$ the upper branch is more populated and thus contains the mode of the distribution, while the lower branch contributes to the $-1.65\sigma$ spread. For [FeH]$>-0.3$, the branches switch and the bifurcation is reflected in $+1.65\sigma$ curve.\footnote{We do not discuss the difference between $+1.65\sigma$ and $-1.65\sigma$ curves for [Al/Fe] since a selection cut is applied to this element restricting the lower wing of the distribution for the in-situ stars}. 

\label{lastpage}

\end{document}